\documentclass[twocolumn]{aastex701}

\usepackage{xcolor}
\usepackage{soul}
\usepackage{romannum}

\shorttitle{Ro-vibrational Spectroscopy of AB Aurigae}
\shortauthors{Kozdon et al.}
\graphicspath{{./}{figures/}}

\begin{document}
\title{${}^{12}$CO Ro-vibrational Spectroscopy of AB Aurigae ---\\ A Potential Point Source is Present}

\author{Janus Kozdon}
\affiliation{Department of Physics and Astronomy, 118 Kinard Laboratory, Clemson University, Clemson, SC 29634-0978, USA}
\email[show]{jkozdon@clemson.edu}  

\author{Jeffrey Fung}
\affiliation{Department of Physics and Astronomy, 118 Kinard Laboratory, Clemson University, Clemson, SC 29634-0978, USA}
\email{fung@clemson.edu}  

\author{Sean D. Brittain}
\affiliation{Department of Physics and Astronomy, 118 Kinard Laboratory, Clemson University, Clemson, SC 29634-0978, USA}
\email{sbritt@clemson.edu}  

\author{Stanley Jensen}
\affiliation{Department of Physics and Astronomy, 118 Kinard Laboratory, Clemson University, Clemson, SC 29634-0978, USA}
\email{skjensenjr@gmail.com}  

\author{Josh Kern}
\affiliation{Department of Physics and Astronomy, 118 Kinard Laboratory, Clemson University, Clemson, SC 29634-0978, USA}
\email{jkern417@gmail.com}  

\author{Cory Padgett}
\affiliation{Department of Physics and Astronomy, 118 Kinard Laboratory, Clemson University, Clemson, SC 29634-0978, USA}
\email{cpadge4@g.clemson.edu}  

\author{Yasuhiro Hasegawa}
\affiliation{Jet Propulsion Laboratory, California Institute of Technology, Pasadena, CA 91109, USA}
\email{yasuhiro.hasegawa@jpl.nasa.gov}  

\begin{abstract}
The Herbig Ae star AB Aurigae hosts a vast, low-inclination protoplanetary disk that exhibits a plethora of substructures, including the protoplanet candidate AB Aur b. We present M-band spectroscopic data taken with NASA IRTF from Feb 2024 covering multiple position angles that captured emission from an off-centered, low temperature, and compact source. Analysis of the ${}^{12}$CO \romannum{5}=1-0 low-J ro-vibrational emission line profiles and spectroastrometric signals localizes the source at around an orbital radius of 65~au and a position angle of 143$\degr$. These coordinates are distinctly different from those of AB Aur b, which was not detected. Although there is no obvious explanation for the detected source, if we assume it is a circumplanetary disk, then its maximum temperature would be about 550~K and its maximum radius would be about 5~au. Our results alludes to a previously unknown companion that may be residing in the AB Aurigae system.
\end{abstract}

\section{Introduction} \label{sec:intro}
AB Aurigae (referred to as AB Aur hereafter) is a  young, $\sim4$~Myr \citep{DeWarf2003}, Herbig Ae star (spectral type of A0Ve \citep{Mooley2013}) that is located 156~pc \citep{Gaia2020} away and is part of the Taurus-Auriga association \citep{Mooley2013}. It has a mass of 2.4~$M_\sun$ \citep{Rodriguez2014} and is surrounded by a gas-rich protoplanetary disk \citep{Pietu2005} that is inclined by 23\degr \citep{Tang2017}. With CO being first detected by \cite{Mannings1997}, gas emission ranges from 0.3~au \citep{diFolco2009} to $\sim1600$~au \citep{Speedie2025, Pietu2005, Corder2005}, this low-inclination protoplanetary disk is ideal for studying material distribution \citep{Bae2022} as well as substructures of disks \citep{Andrews2020}.

Ever since the first images that resolved millimeter emission \citep{Mannings1997} and scattered light \citep{Grady1999}, many disk features/substructures have been identified in the AB Aur disk. Some of the more notable ones include a dust ring centered around $\approx$100~au \citep{Riviere-Marichalar2024} extending to $\approx$200~au \citep{Tang2012}. The western portion of the dust ring is brighter \citep{Tang2012} where there might be a dust trap associated with a decaying vortex \citep{Fuente2017}. 
Gaseous spirals are present within the dust ring \citep{Tang2017} (see also \cite{Beck2019}) and beyond it \citep{Lin2006, Corder2005} beyond it. The outer spiral shows evidence of interactions with further outer disk features \citep{Dutry2024} instigated by a possible encounter with a dense cloudlet \citep{Kuffmeier2020, Nakajima1995}.

Multiple localized sources have been proposed to be present in AB Aur's protoplanetary disk. Two are point sources imaged by \cite{Boccaletti2020} where one is located almost 30~au southwest of the star, while the other is $\sim110$~au northward. \cite{Currie2022} identified another localized source, not a point source, about 94~au south of the star. These potential planetary companions may help explain the development of disk features such as the inner spiral \citep{Poblete2020, Tang2017} and the apparent highly perturbed nature of the cavity \citep{Calcino2024}. However, it is debated if the companion proposed by \cite{Currie2022} is truly an embedded protoplanet \citep{Bowler2025, Currie2024} or a disk feature \citep{Biddle2024, Zhou2023}. Recently, \cite{Bowler2025} recovered the \cite{Currie2022} source from its H$_{\alpha}$ emission but not the sources from \cite{Boccaletti2020}.

This study aims to capture the high-resolution ro-vibrational CO emission from the AB Aur system to constrain properties like the distribution and thermophysical characteristics of its protoplanetary disk. Ro-vibrational CO emission of this source was first detected by \cite{Brittain2002} (see also \cite{Blake2004}) where it was shown that the gas is thermally excited. While the emission lines were not spectrally resolved in these early studies, the availability of instrumentation with higher spectral resolution has enabled more detailed analysis of these lines. Like with \cite{Jensen2024} who found that the gas was not consistent with an axisymmetric distribution of emission around the star. Similar spectroscopic analysis of these transitions have previously constrained circumstellar disk properties such as eccentricity \citep{Liskowsky2012} and continuity \citep{Kozdon2023}.

Spectroastrometric (SA) signals are the relative flux-weighted offsets from the continuum and can be used to determine the spatial locations of protoplanetary disk emission \citep{Pontoppidan2008}. 
See \cite{Brittain2015} and references herein for an overview of spectroastrometry.
When analyzed in conjunction with the line profiles, the SA signals can elucidate various non-axisymmetric and non-Keplerian structures/components. These include circumplanetary disks \citep{Brittain2019} or disk winds \citep{Pontoppidan2011} (for another application see \cite{Mendigutia2018}). Based on the SA signal from AB Aur, \cite{Jensen2024} concluded that an additional, non-axisymmetric source is present somewhere in the southern portion of the protoplanetary disk. However, the irregular seeing conditions and limited position angle coverage of their observations hampered the interpretation of their results. Here we present spectral and spectroastrometric analysis (similar to \cite{Kozdon2023} and \cite{Jensen2024}) of further observations of AB Aur to characterize the emission and constrain the location of the source.

The following section, Sec. \ref{sec:OBS}, overviews the data collection routine and reduction process. In Sec. \ref{sec:RESULTS}, the average spectrum is presented where the high and low energy transitions are discussed further in Sec. \ref{sec:high_vs_low}. The hydrogen Pf$\beta$ transition is discussed in Sec. \ref{sec:hydrogen_transition}. Detailed analysis of the ${}^{12}$CO line profiles and SA signals are performed in Sec. \ref{sec:LINES_FIT} where individual attention is given to the high energy (Sec. \ref{sec:LINES_HI}) and low energy (Sec. \ref{sec:LINES_LO} and Sec. \ref{sec:LINES_LOCPD}) transitions. The results from Sec. \ref{sec:LINES_FIT} inform a hydrodynamic simulation in Sec. \ref{sec:HYDRO} for hypothetical testing of possible disk features that may develop due to a planetary companion. Lingering questions are addressed in Sec. \ref{sec:DISS} and discussion pertaining to AB Aur b specifically is reserved for Sec. \ref{sec:notABAurb}. Lastly, Sec. \ref{sec:CONC} highlights our main results (see also \citep{Kozdonthesis}).

\section{Observations and Reductions}
\label{sec:OBS}
The data was collected with the iSHELL cross-dispersion echelle spectrograph at the NASA Infrared Telescope Facility (IRTF) \citep{Rayner2022} for three epochs in 2024. While observing, an ABBA nodding pattern, with no chopping, was used to remove sky emission. For the science target, the images were taken with 5 coadds of 10 second exposures whereas the images of the standard star used 6 coadds of 10 second exposures. The number of ABBA nodding cycles and the approximate integration times are presented in Table \ref{tab:epochs}. For the standard star (HR 2088) the number of cycles from each epoch are 3, 4 and 5.

Depending on the seeing conditions, we either adopted a slit width of 0.375$\arcsec$ or 0.75$\arcsec$ (see Table \ref{tab:epochs}). The 0.375$\arcsec$ wide slit provides a resolving power of R$\sim92,000$ and a spectral resolution of $\sim3.3$~km~s$^{-1}$ whereas the 0.75$\arcsec$ provides R$\sim60,000$ and $\sim5.0$~km~s$^{-1}$ \citep{Banzatti2022}. For this study, we adopted spectral resolutions of 3.5~km~s$^{-1}$ and 6.0~km s$^{-1}$. More details of the instrument and its performance can be found in \cite{Rayner2022}.

\begin{figure}[t]
    \centering
    \includegraphics[clip, trim = 0.1cm 0.1cm 0.1cm 0.1cm, width=0.99\linewidth]{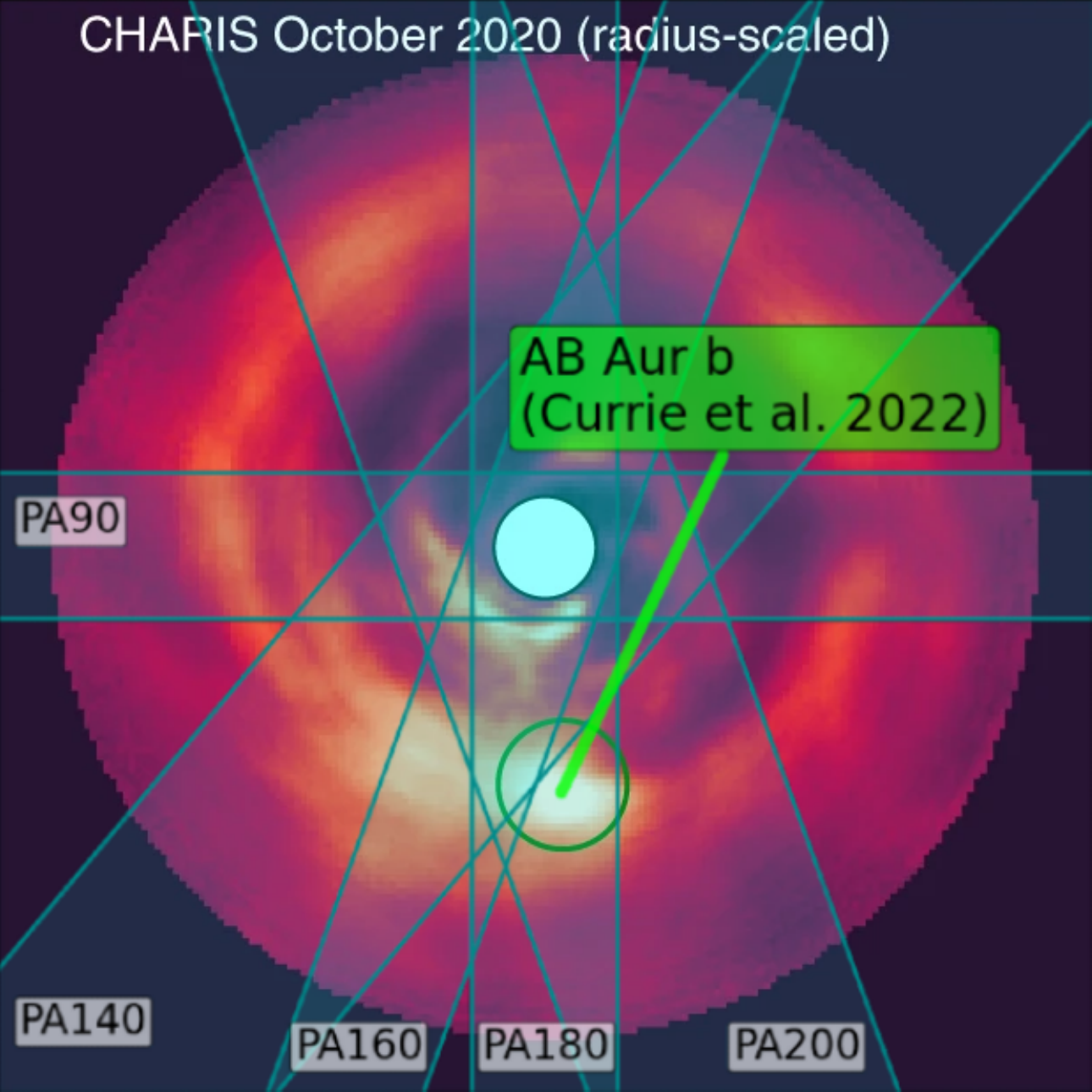}
    \caption{Schematic of the slit observations (teal) presented in Table \ref{tab:epochs} (Sec. \ref{sec:OBS}). The overlay uses the CHARIS image from \cite{Currie2022} to illustrate the attained coverage. This slit arrangement allows for the analysis of emission from outer disk sources (Sec. \ref{sec:LINES_LO}), like the protoplanet candidate AB Aur b.}
    \label{fig:Observations}
\end{figure}

Throughout each epoch, the observing slit was oriented along multiple position angles (PAs) to capture varying amounts of non-axisymmetric emission from the AB Aur system --- like from AB Aur b. For the most part, every PA collection was followed by an anti-parallel one with the same integration settings --- required for measuring the SA signals \citep{Pontoppidan2008}. The captured PA pairs are 180\degr/0\degr, 90\degr/270\degr, 160\degr/340\degr and 140\degr/320\degr. Data was also captured along 200\degr but not its anti-parallel (20\degr) and, as such, is neglected from the majority of the analysis. See Fig. \ref{fig:Observations} for a showcase of the acquired coverage of the system for this study. Data taken along the system's semi-major (60\degr/240\degr) and semi-minor axis (150\degr/330\degr) are presented in \cite{Jensen2024} along with an additional PA pair of 24\degr/204\degr. However, their data is not incorporated into our analysis.

The data reduction was done in a usual manner for IR spectroscopy (flat fielding, image stacking, spectra extraction, etc.) using custom routines, highlighted in \cite{Brittain2018}, that utilizes a sky model generated by Sky Synthesis Program \citep{Kunde1974}. Because iSHELL has an auto guider the ABBA set images did not need to be shifted for self-consistency. The SA signals are collected by fitting a skewed Moffat function to the continuum's point spread function and measuring the peak counts' location. The SA measurements as well as the data reduction process are performed the same way as \cite{Jensen2024} (see also \cite{Jensen2021}).

\begin{deluxetable*}{ccccccc}[t]
\tablecaption{Observation log \label{tab:epochs}}
\tablewidth{0pt}
\tablehead{
Epoch & UT & PA & cycles & integration time & seeing & slit width \\
\# & \_ & [deg] & \# & [min] & [arcsec] & [arcsec] }
\startdata
1 & 11 Feb 2024 & 180/0   & 8/8 & 19/19 & 0.6 & 0.375 \\
  &             & 90/270  & 8/8 & 19/19 & 0.8 & 0.375 \\
2 & 12 Feb 2024 & 160/340 & 16/16 & 38/38 & 0.9 & 0.375 \\
3 & 17 Feb 2024 & 140/320 & 14/14 & 34/34 & 1.2 & 0.750 \\
  &             & 200     & 12 & 29 & 1.2 & 0.750 \\
\enddata
\end{deluxetable*}

The average spectrum of the PA's captured with the 0.375$\arcsec$ are presented in Fig. \ref{fig:Spectrum}. The CO transitions of interest (elaborated below) are labeled in red. Analysis is done on the transitions that are not blended with other features, like CO \romannum{5} = 2-1 ro-vibrational states, and are not too faint or have significant telluric corruption. Lastly, the transitions for analysis must be well behaved between all the PAs captured to keep the comparisons straightforward. The chosen transitions ended up being the P(27), P(26), P(25), P(19), P(18), P(8), P(3) and R(2) transitions and they are labeled (red) in Fig. \ref{fig:Spectrum}.

The measured fluxes of the chosen transitions, $\mathfrak{F}_{\rm J}$ (where J describes the lower rotational state), are calculated using the AllWISE Multiepoch Photometry Database value centered on 4.6~$\mu$m and multiplying it with the measured equivalent widths. The equivalent widths are measured via numerical summation and the error bars are calculated from usual error propagation rules (see \cite{Vollmann2006}). The $\mathfrak{F}_{\rm J}$ values (with units of W~m$^{-2}$) are plotted in Fig. \ref{fig:rotational_temperature} according to the following equation:
\begin{equation}
    \frac{k}{hcB}\ln\frac{\mathfrak{F}_{\rm J}}{\tilde{\nu}_{\rm J}^{4} ({\rm J+J^{\prime}+1})} = \frac{-\rm J{^{\prime}}(J{^{\prime}}+1)}{T_{\rm rot}} + {B_{\rm inter}} ,\
    \label{eq:rotaiton_temperature}
\end{equation}
where $k_{\rm{B}}$ is the Boltzmann constant, $h$ is the Planck constant, $c$ is the speed of light and $B$ is a rotational constant. $\tilde{\nu}_{\rm J}$ is the central wavenumber and J$^\prime$ describes the upper state. $T_{\rm rot}$ is the rotational temperature and can be understood as the system's average weighted temperature. $B_{\rm inter}$ is a systemic offset that depends on multiple constants including the earlier mentioned ones but also the acquired solid angle, a spectroscopic constant and the partition function (see \cite{Troutman2011, Brittain2003}. Also, Eqn. \ref{eq:rotaiton_temperature} assumes optically thin emission where no absorption of the background continuum occurs.

To enhance the signal-to-noise ratio for spectroscopic analysis, the chosen transitions are stacked together, without renormalization, to create an average profile/signal. Velocity corrections are performed for the local standard of rest, the system velocity and the individual epoch's relative motion. Two line profiles and two SA signals are constructed for most epochs --- one for the high-J and low-J transitions. The high-J transitions are defined as those with J$\geq$20, or $\tilde\nu\leq$ 2060~cm$^{-1}$, and includes the P(25), P(26) and P(27) transitions whereas the low-J transitions, or J$<$20 and $\tilde\nu>$ 2060~cm$^{-1}$, includes the R(2), P(3), P(8), P(18) and P(19) transitions. The average stacked emission line profiles and SA signals of the ${}^{12}$CO lines from each PA are presented in Fig. \ref{fig:ABAUR_all} with the associated equivalent widths presented in Table \ref{tab:EQW}.

Also captured in our data is the hydrogen Pf$\beta$ transition (blue; Fig. \ref{fig:Spectrum}). Its profiles are presented in Fig. \ref{fig:hydrogen} with the associated equivalent widths in Table \ref{tab:EQW}. Because it is an individual transition, no stacking was performed and no spectroastrometric information can be extracted. Centered on 2148.5 cm$^{-1}$ this transition is superimposed with certain ${}^{12}$CO \romannum{5}=1-0 and \romannum{5}=2-1 lines. These features, along with the tellurically corrupted segments, are masked via linear interpolation (dashed lines; Fig. \ref{fig:hydrogen}).

\section{Results}
\label{sec:RESULTS}
\begin{figure*}[t]
    \centering
    \includegraphics[clip, trim = 4.5cm 4.0cm 5.5cm 4.5cm, width=0.99\textwidth]{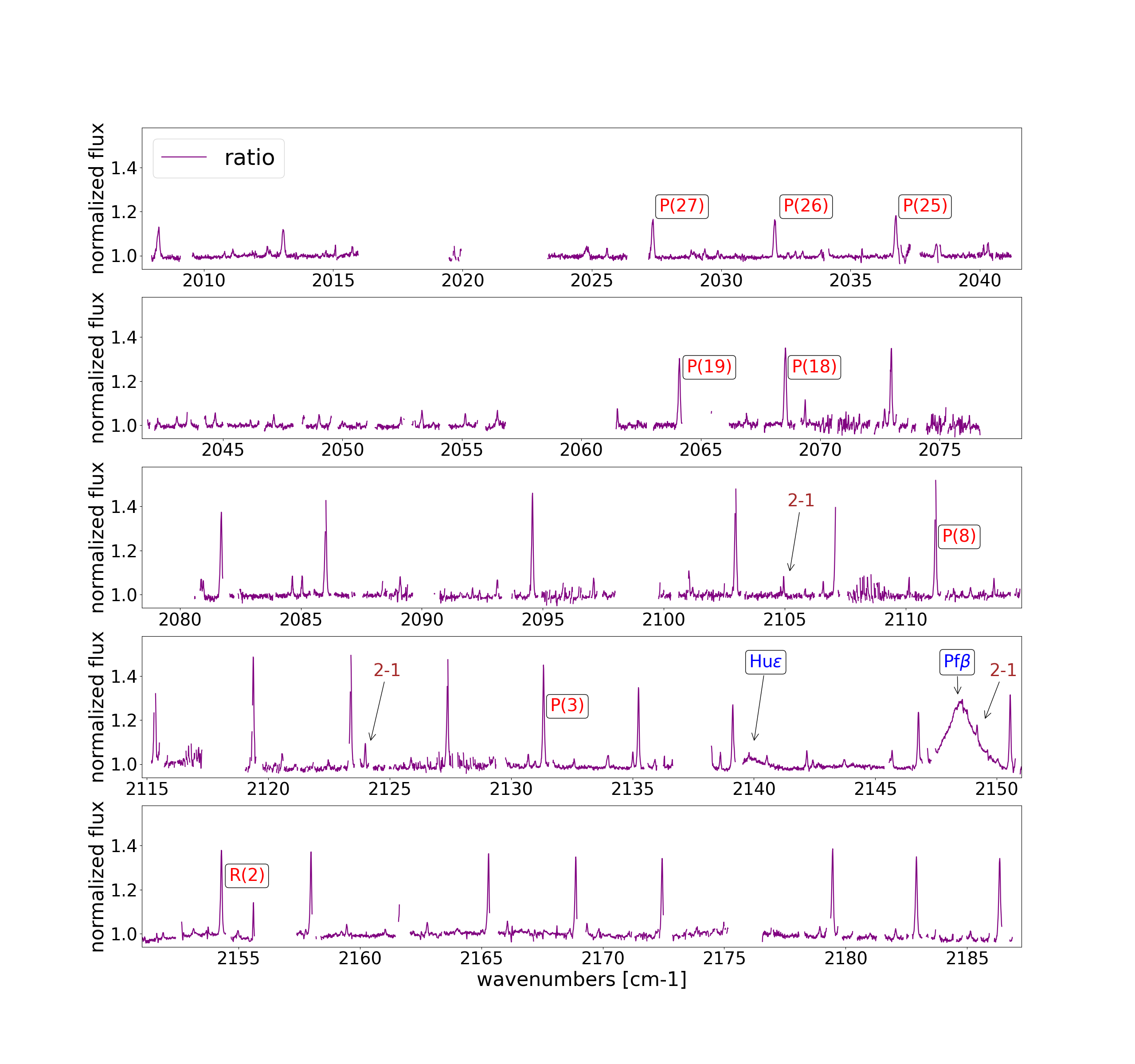}
    \caption{The average spectrum of the observations (Table \ref{tab:epochs}; Sec. \ref{sec:OBS}). The CO \romannum{5}=1-0 lines (red) and the hydrogen Pf$\beta$ transition (blue) are analyzed in Sec. \ref{sec:RESULTS}). A select few CO \romannum{5}=2-1 lines (brown) are labeled and their presence indicates a high temperature (Sec. \ref{sec:LINES_HI}). Also pointed out is the Hydrogen Hu$\epsilon$ transition.}
    \label{fig:Spectrum}
\end{figure*}

\begin{figure}[t]
    \centering
    \includegraphics[clip, trim = 2.0cm 0.2cm 7.2cm 3.0cm, width=0.99\linewidth]{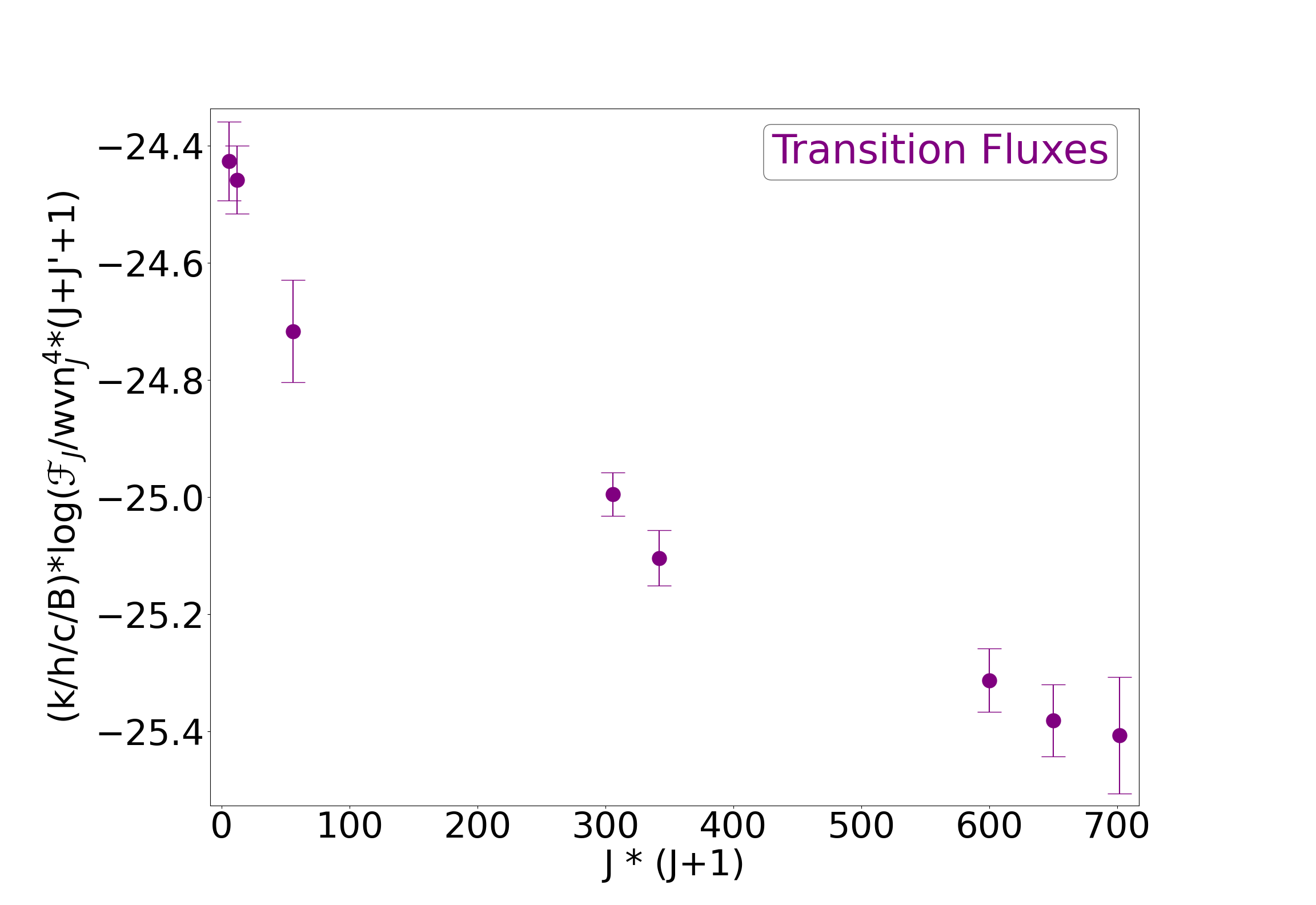}
    \caption{Rotational diagram for the transitions labeled in Fig. \ref{fig:Spectrum}. In order the data points are R(2), P(3), P(8), P(18), P(19), P(25), P(26) and P(27). If the observed fluxes originates from an optically thin, isothermal source then the data points should follow a line, which it doesn't (Sec. \ref{sec:RESULTS}). This is an indication that the emission from AB Aur's protoplanetary disk has a gradient in optical depth and temperature.}
    \label{fig:rotational_temperature}
\end{figure}

The average spectrum of AB Aur is presented in Fig. \ref{fig:Spectrum} where the CO \romannum{5} = 1-0 transitions that were selected for analysis (red) are labeled. The rotational diagram of the those transitions are displayed in Fig. \ref{fig:rotational_temperature}. The PA average line profiles and spectroastrometric signal of the high-J and low-J transitions are presented in Fig. \ref{fig:ABAUR_all}. Features and characteristics of these profiles/signals are discussed in Sec. \ref{sec:high_vs_low} and rigorous analysis is done in Sec. \ref{sec:LINES_FIT}. The line profiles for the Hydrogen Pf$\beta$ transition (blue) are illustrated in Fig. \ref{fig:hydrogen} and discussed in Sec. \ref{sec:hydrogen_transition}. See Table \ref{tab:EQW} for the equivalent width measurements.

The M-band spans 1950 to 2200 cm$^{-1}$ and covers the ${}^{12}$CO \romannum{5} = 1-0 ro-vibrational transitions from R(15) to P(42). Fig. \ref{fig:Spectrum} is centered around the region of the spectrum where data is captured. Along with the CO \romannum{5} = 1-0 transitions are the \romannum{5} = 2-1 transitions where a select few are pointed out (brown). The CO \romannum{5} = 3-2 transitions may also be present but are not as easily discernible. Multiple CO bands being populated can have implications for the degree of UV fluorescence \citep{Brittain2009, Krotkov1980} in the system. Also, \cite{Brittain2003} did not observe the \romannum{5} = 2-1 transitions, which may suggest that the system has increased in temperature since then.

The system's rotational temperature can be determined by performing a linear fit to Fig. \ref{fig:rotational_temperature} or Eqn. \ref{eq:rotaiton_temperature} (see \cite{Troutman2011, Brittain2003}). Even though our derived rotational temperature is $T_{\rm rot}$ $\approx$ 1000~K, we believe this to be inaccurate as the data points are not linear. \cite{Brittain2003} studied the same transitions and captured the same non-linear feature where they concluded that two different temperature components were required between the low-J and high-J transitions. However, we believe the same phenomenon can result from temperature gradients and optical depth differences. Our analysis of the line profiles (Sec. \ref{sec:LINES_FIT}) accommodates these effects.

\subsection{High-J Lines vs. Low-J Lines}
\label{sec:high_vs_low}
\begin{figure*}[t]
    \centering
    \includegraphics[clip, trim = 4.2cm 2.4cm 6.4cm 5.2cm, width=0.99\textwidth]{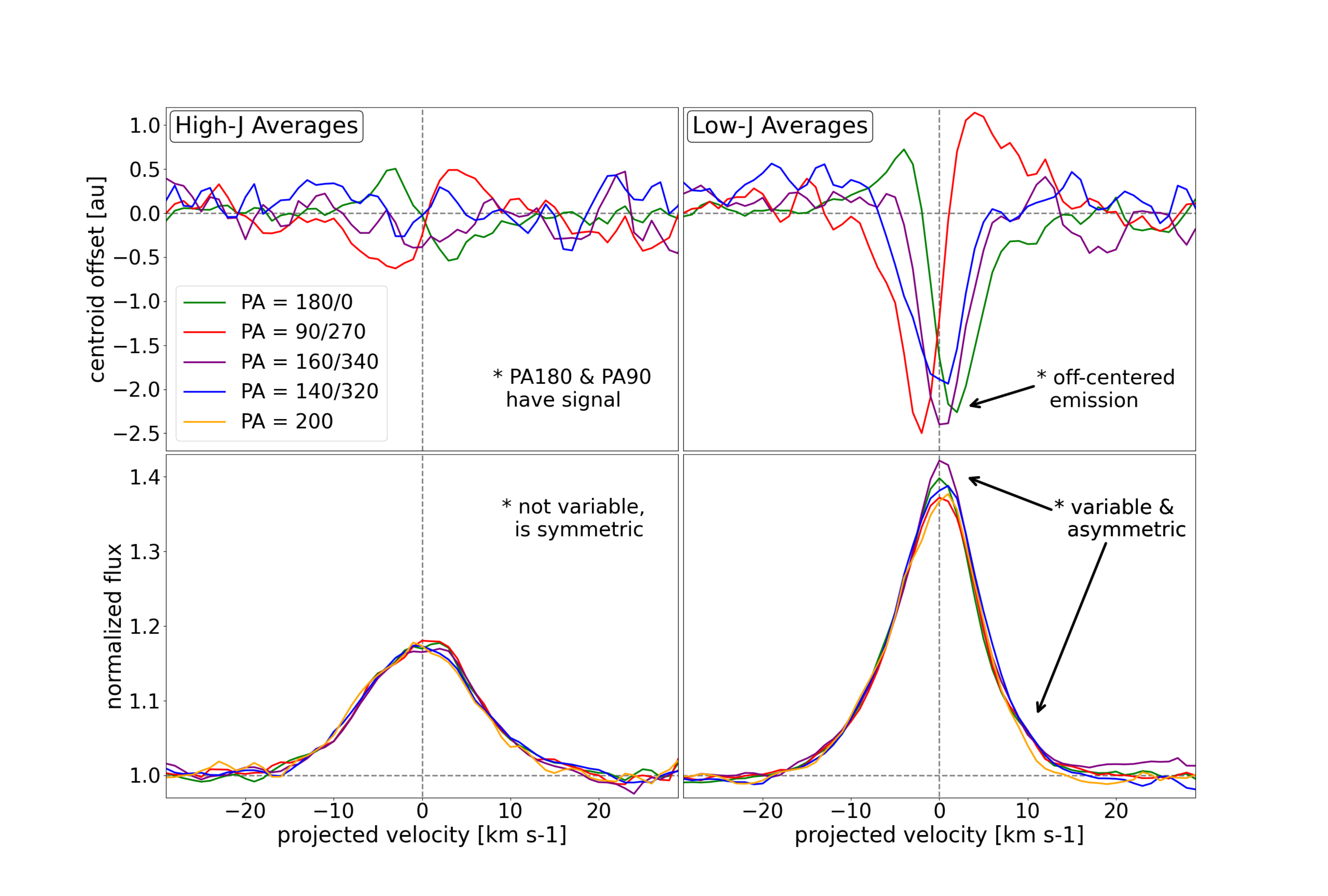}
    \caption{The PA averages of the emission line profiles (bottom) and spectroastrometric signals (top) from AB Aur's high-J (left) and low-J (right) ${}^{12}$CO ro-vibrational \romannum{5} = 1-0 transitions (Sec. \ref{sec:high_vs_low}). Variable emission between PAs was captured in the low-J line profiles as well as a spatial offset in the spectroastrometric signals. This behaviour is not mirrored in the high-J equivalent and that can be an indication of a low-temperature substructure being present in the protoplanetary disk (Sec. \ref{sec:LINES_LO} and Sec. \ref{sec:LINES_LOCPD}). Also, the high-J line profiles are used for determining the thermophysical environment of the protoplanetary disk's emitting surface layer (Sec. \ref{sec:LINES_HI}).}
    \label{fig:ABAUR_all}
\end{figure*}

The ${}^{12}$CO high-J average line profiles (bottom left: Fig. \ref{fig:ABAUR_all}) all appear symmetric and exhibit no clear differences between PAs. The associated SA signals (top left) are only discernible in PA180 and PA90. The shape of these SA signals, alongside the line profiles' left-right symmetry, indicate that the high-J emission is axisymmetric. Therefore, these profiles, not SA signals, are used to constrain the emitting layer's temperature and surface density conditions in Sec. \ref{sec:LINES_HI}.

Contrary to the high-J line profiles, the low-J counterparts (bottom right: Fig. \ref{fig:ABAUR_all}) did noticeably vary between PAs --- particularly PA160. It may be due to a potential substructure in the protoplanetary disk whose emission is being captured to varying degrees. The accompanying SA signals (top right) corroborate this theory as they captured a significant offset in a common direction. These line profiles, including SA signals, are analyzed in Sec. \ref{sec:LINES_LO} and Sec. \ref{sec:LINES_LOCPD} to constrain the intensity and location of the potential substructure.

The low-J line profiles may also be asymmetric. Consistently, the blue-shifted wings (negative projected velocities) appear to extend a few km~s$^{-1}$ farther than the red-shifted wings (positive projected velocities). \cite{Jensen2024} did not observe this asymmetry, but their line profiles were constructed by stacking both high-J and low-J lines, which could have muddied this behavior. Whatever disk feature is responsible for this high-velocity asymmetry ($\pm$15~km~s$^{-1}$) is separate from the one responsible for the emissions captured at near-zero velocities. Even though the degree of asymmetry is minor, it is not mirrored in the high-J counterparts which; which suggests the feature responsible has a low temperature. This work will focus on discerning the origin of the line profile behavior and SA signal offsets, whereas studies of the high-velocity asymmetry will be deferred.

\begin{deluxetable}{ccccc}[b]
\tablecaption{PA Average Equivalent Widths
\label{tab:EQW}}
\tablewidth{0pt}
\tablehead{
\colhead{Epoch} & \colhead{PA} & \colhead{CO high-J} & \colhead{CO low-J} & \colhead{H Pf$\beta$} \\
\colhead{\#} & \colhead{\degr} & \colhead{km s$^{-1}$} & \colhead{km s$^{-1}$} & \colhead{km s$^{-1}$}}
\startdata
1 & 180 & 2.9$\pm$0.3 & 4.8$\pm$0.5 & 49.7$\pm$5.1 \\
  & 90  & 2.9$\pm$0.4 & 4.7$\pm$0.5 & --- \\
2 & 160 & 2.8$\pm$0.3 & 5.0$\pm$0.4 & 38.0$\pm$5.9 \\
3 & 140 & 2.8$\pm$0.4 & 4.9$\pm$0.5 & 49.8$\pm$5.1 \\
  & 200 & 2.8$\pm$0.4 & 4.6$\pm$0.6 & --- \\ 
\hline
  & avg & 2.8$\pm$0.4 & 4.8$\pm$0.5 & 46.1$\pm$5.4 \\
\enddata
\centering
\tablecomments{The equivalent widths of the PA average transitions presented in Fig. \ref{fig:ABAUR_all} and the hydrogen transitions in Fig. \ref{fig:hydrogen}.}
\end{deluxetable}

The hydrogen transition (Fig. \ref{fig:hydrogen}) can be used to gauge stellar variability. The hydrogen equivalent widths (Table \ref{tab:EQW}) are all measured to be within 2$\sigma$ of the average, and, as such, no significant stellar variability was found during our observation run. Further discussion is reserved for Sec. \ref{sec:hydrogen_transition}.

\subsection{Hydrogen Line}
\label{sec:hydrogen_transition}
\begin{figure}[t]
    \centering
    \includegraphics[clip, trim = 0.6cm 0.4cm 3.2cm 2.9cm, width=0.99\linewidth]{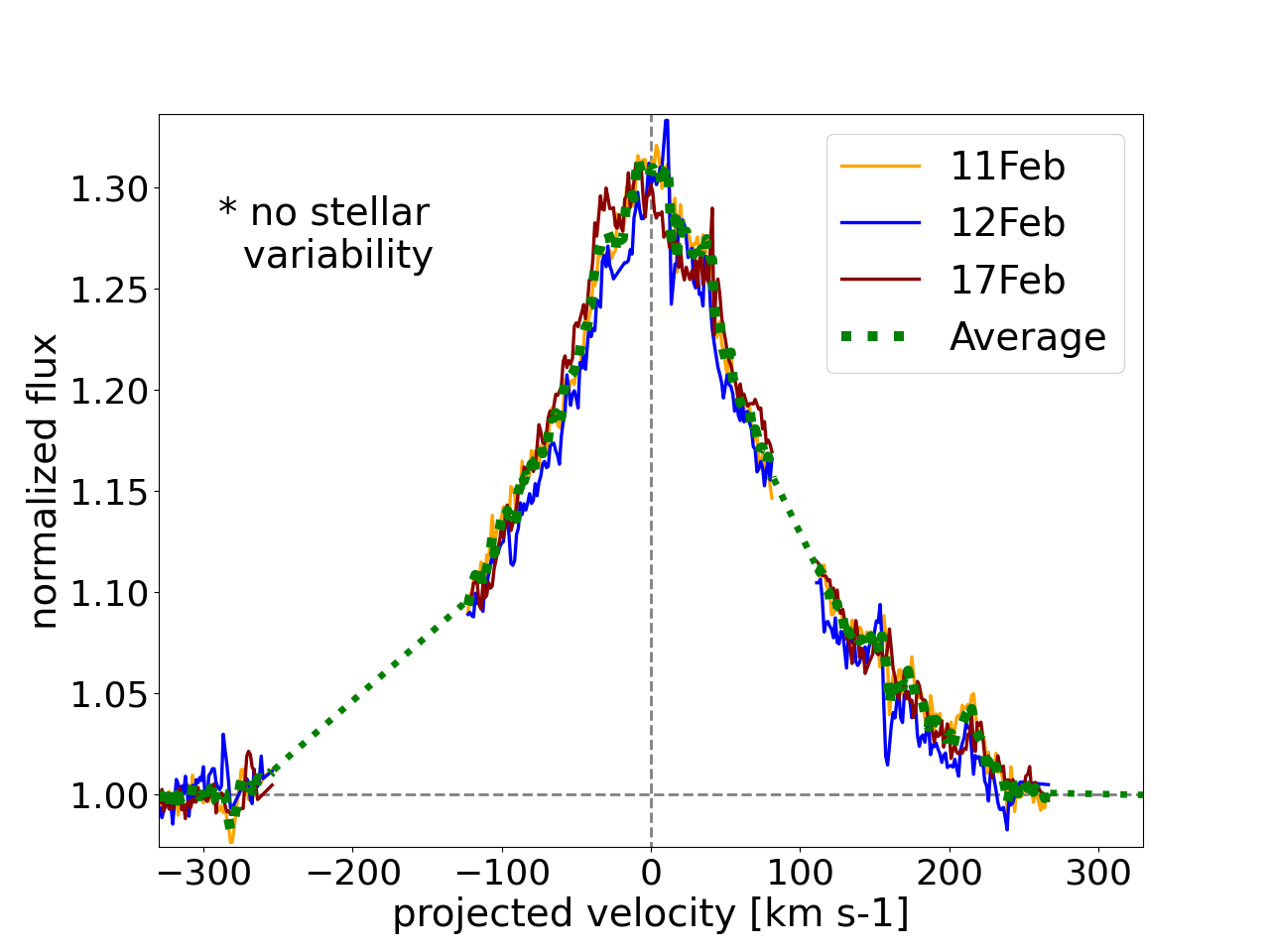}
    \caption{The Hydrogen Pf$\beta$ transition of each observing epoch (Table \ref{tab:epochs}; Sec. \ref{sec:RESULTS}). No stellar variability was detected during the 1 week span of observations. The average profile (green) is utilized for determining the velocity spread and calculating a stellar accretion rate in Sec. \ref{sec:hydrogen_transition}. The dashed lines are where external features, like telluric corruption and CO lines, are present and were masked over via linear interpolation.}
    \label{fig:hydrogen}
\end{figure}

Fig. \ref{fig:hydrogen} displays the hydrogen Pf$\beta$ transition from each observing epoch. Because no significant changes were detected, calculations are performed on the average profile (green). Fitting the average profile with a Gaussian function gives a full-width-at-half-maximum of FWHM$_{\rm Pf\beta}$ = 202~km~s$^{-1}$ (Gaussian standard deviation of $\sigma_{\rm Pf\beta}$ = 86~km~s$^{-1}$). \cite{Brittain2003} studied the same transition from 2001/2002 and reported a value of FWHM$_{\rm Pf\beta} \sim130$~km~s$^{-1}$ ($\sigma_{\rm Pf\beta} \sim 55$~km~s$^{-1}$). This corresponds to an $\approx$55\% increased in the velocity broadening over the 23~years between their study and ours. Presuming this emission originates from the host star, and is not extended relative to the continuum, this change can be mostly attributed to changes in outflow broadening \citep{Kuiper2016} from stellar winds. However, it can also involve a combination of other macro-broadening effects like stellar rotation \citep{Collins1995}, granulation \citep{Guo2022} or starspots \citep{Dorval2024}.

During our observation run, the average equivalent width of the hydrogen Pf$\beta$ transition is 46.1$\pm$5.4 km s$^{-1}$. \cite{Salyk2013} studied the same transition back in 2001 and reported a value of $\approx$44 km s$^{-1}$. Because their value falls within our uncertainty, we report no significant change in the hydrogen emission over the 23~year period. \cite{Brittain2003} mentioned that their equivalent widths changed by $\sim80\%$ between their 2001 and 2002 observations, though they did not provide the measurements themselves. These point to potential significant flux changes within yearly timescales. However, the behavior of the variability must be longer than a week because the Pf$\beta$ transition remained consistent during our observations (Fig. \ref{fig:hydrogen}; Table \ref{tab:EQW}).

We infer AB Aur's stellar accretion rate using the hydrogen Pf$\beta$ luminosity in the same manner as \cite{Kozdon2023} did for CI Tau by using the empirical relation provided in \cite{Salyk2013} (see also \cite{Marleau2022}. Using a distance of $d$ = 156~pc \citep{Gaia2020}, a stellar radius of 2.5 $R_{\sun}$ \citep{Li16} and a stellar mass of 2.4~$M_\sun$ \citep{Rodriguez2014} the average accretion rate during our observing run was $\dot{\rm M} = $1.6$^{+1.9}_{-0.9}\times10^{-7} M_{\sun}$ yr$^{-1}$. Our derived accretion rate for AB Aur is not atypical for Herbig Ae/Be stars (see \cite{Wichittanakom2020}). Note that the large error bars are mainly due to uncertainty in the empirical relations and that the error in our Pf$\beta$ luminosity is smaller ($\approx$ 12\%).

It is expected that hydrogen emission originates not only from the host star but also from accreting planetary companions \citep{Eriksson2020}. Due to our signal-to-noise, we did not detect a feature in the Hydrogen line profile (Fig. \ref{fig:hydrogen}) that could be attributed to sources other than the host star. As such, our hydrogen data does not provide information pertaining to a companion. However, this system's previously reported planetary candidates have been detected through their gas emissions \citep{Boccaletti2020, Currie2022}. 

\subsection{Emission Line Profile Analysis}
\label{sec:LINES_FIT}
Separate fitting routines were performed on AB Aur's ${}^{12}$CO profiles/signals; one for the high-J lines (left; Fig. \ref{fig:ABAUR_all}) to constrain the thermophysical environment (Sec. \ref{sec:LINES_HI}) and another for the low-J lines (right; Fig. \ref{fig:ABAUR_all}) to characterize the potential source (Sec. \ref{sec:LINES_LO}). To optimize computational resources, we employ a multi-grid method --- see App. \ref{sec:Appendix} for a description. The benefit of this methodology is that it provides increased resolution for the high-velocity inner portions of the disk.

\subsubsection{Protoplanetary Disk Model (High-J Fit)}
\label{sec:LINES_HI}
Our analysis of the ${}^{12}$CO high-J ro-vibrational stacked emission line profiles of AB Aur (left; Fig. \ref{fig:ABAUR_all}) begins by assuming a circular two-dimensional disk with the inner and outer radial bounds of $r_{\rm in}$ and $r_{\rm out}$, respectively. Simple power laws are adopted for a radially dependent temperature, $T_{\rm rot}(r)$, and surface number density, $N(r)$:
\begin{equation}
    T_{\rm rot}(r)=T_0\left(\frac{r}{r_{\rm{in}}}\right)^{-T_\alpha} \,, \quad N(r)=N_0\left(\frac{r}{r_{\rm{in}}}\right)^{-N_\alpha} \,.
    \label{eq:power_laws}
\end{equation}
Here, $T_{0}$ and $N_{0}$ are the temperature and surface density values normalized at the inner radii where $T_{\alpha}$ and $N_{\alpha}$, the exponents of the power laws, describes the rate of change of the values. For simplicity, we assume local thermodynamic equilibrium (LTE), which allows us to approximate the rotational temperature as the vibrational temperature. We caution the reader that this assumption may not be valid in all regions of the disk. Together with $r_{\rm in}$ and $r_{\rm out}$, this model has six parameters in total that are fitted.

The state populations down to the disk's photosphere, $N_{\rm J}$ ($\rm J$ being the lower rotational state), are calculated with the Boltzmann equation:
\begin{equation}
    N_{\rm J}= \frac{N(r) g_{\rm J}}{ Q(T)} \exp\left(\frac{- E_{\rm J}}{ k_{\rm B} T_{\rm rot}(r)}\right) \,.
    \label{eq:NJ}
\end{equation}
Here, $g_{\rm J}$ and $E_{\rm J}$ are the degeneracies and excitation energies of state J, respectively. $Q(T)$ is the temperature dependent partition function for ${}^{12}$C${}^{16}$O and $k_{\rm B}$ is the Boltzmann constant. The central optical depth of the transition to state J, $\tau_{\rm J}$, is calculated with
\begin{equation}
    \tau_{\rm J}=\frac{N_{\rm J}}{ 8\pi^{\frac{3}{2}}b_{\rm therm}(T_{\rm rot})}\frac{g_{\rm J^{\prime}} A_{\rm J}}{ g_{\rm J}\tilde{\nu}_{\rm J}^3} \,,
    \label{eq:tau}
\end{equation}
where $A_{\rm J}$ and $\tilde{\nu}_{\rm J}$ are the Einstein $A$ coefficient and wavenumber of state J, respectively. Also, $g_{\rm J^{\prime}}$ is the degeneracy of the upper state. $b$ is the thermal velocity spread and is calculated with
\begin{equation}
    b_{\rm therm}(T_{\rm rot}) = \sqrt{\frac{2k_{\rm B}T_{\rm rot}}{m_{\rm mol}}} \,,
    \label{eq:bspread}
\end{equation}
where $m_{\rm mol}$ is the molecular mass. Finally, the flux densities, or the fluxes evaluated at line centers, of state J, $F_{\rm J}$, are calculated with
\begin{equation}
    F_{\rm J} = (1-e^{-\tau_{\rm J}})\,B_{\rm J}(\tilde{\nu}_{\rm J},T_{\rm rot}) \,,
    \label{eq:transitional_flux}
\end{equation}
where $B_{\rm J}(\tilde{\nu}_{\rm J},T_{\rm rot})$ is the Planck function expressed in units of wavenumbers ($\rm cm^{-1}$). 
We will use this line-center flux density to scale the contributions of flux from different parts of the disk. The microphysics of line broadening due to thermal effects and gas turbulence are not modeled because we expect that our spectral resolution (about 3.5 km s$^{-1}$) is large enough to mask these effects. We have also ignored continuum absorption because the continuum flux density is much lower than $B_{\rm J^{\prime}}(\tilde{\nu}_{\rm J^{\prime}},T_{\rm rot})$. 
We also note that the AB Aur disk contains a large dust cavity about 70 au wide \citep{Pietu2005,Tang2012,Pacheco2016}, which suggests continuum absorption in the inner part of the disk should be negligible. Additionally, dust settling could reduce continuum absorption.
%

The flux densities of the disk are sorted based on their projected velocities, $V_{\rm p}$, which are calculated with
\begin{equation}
    V_{\rm p}(r,\theta) = \sqrt{\frac{GM_{*}}{r}} \sin\theta \sin i \,,
    \label{eq:projected_velocity}
\end{equation}
where $\theta$ is the true anomaly, $i$ is the system inclination and $GM_{*}$ is the host star's standard gravitational parameter. Performing Eqns. \ref{eq:power_laws}-\ref{eq:projected_velocity} throughout the entire disk spatially and spectrally results in a data cube.

The data cube is spatially convolved with 2-dimensional Gaussian kernels that mimic seeing conditions. Also, 2-dimensional arrays resembling the observing slits are applied over the data cube. Thus resulting in multiple data cubes that reflect the observing conditions of each epoch (Table \ref{tab:epochs}).

\begin{figure*}[ht]
    \centering
    \includegraphics[clip, trim = 14.5cm 6.0cm 16.5cm 11.0cm, width=0.99\textwidth]{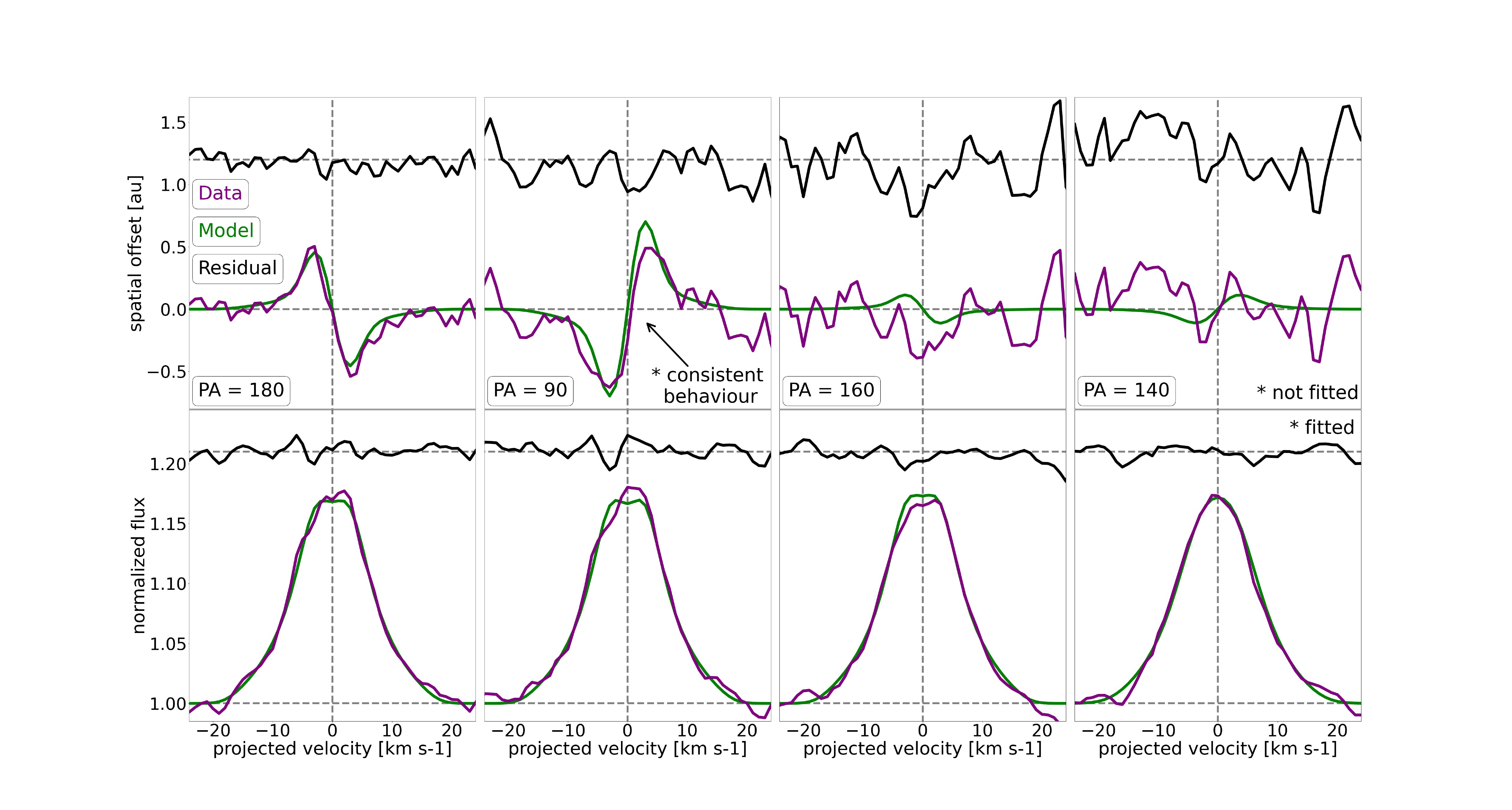}
    \caption{The high-J average transitions (purple) from Fig. \ref{fig:ABAUR_all} with the best-fit results (green). The analysis of these line profiles (not spectroastrometric signals) are described in Sec. \ref{sec:LINES_HI} where the protoplanetary disk's radial extant, surface temperatures and surface densities are constrained (Table \ref{tab:BESTFIT}).}
    \label{fig:bestfit_hiJ}
\end{figure*}

Convolutions are also applied spectrally with a Gaussian profile to accommodate broadening effects. The Gaussian profile is generally described as
\begin{equation}
    G(V-V_{\rm p}) = \frac{1}{b_{\rm Gauss}\sqrt{\pi}} \exp\left(\frac{-(V-V_P)^2}{b^2_{\rm Gauss}}\right) \,,
    \label{eq:gaussian}
\end{equation}
where $b_{\rm Gauss}$ is the overall broadening term and is generally defined as $b_{\rm Gauss}$ = $\sqrt{2}\sigma_{\rm Gauss}$ with $\sigma_{\rm Gauss}$ being the Gaussian standard deviation. $\sigma_{\rm Gauss}$ accounts for multiple broadening effects such as that associated with the instrument profile, $\sigma_{\rm inst}$, and other sources, $\sigma_{\rm other}$, are collectively accounted for by setting $\sigma_{\rm other} = $2~km~s$^{-1}$. The overall Gaussian standard deviation is calculated by $\sigma_{\rm Gauss}$ = $\sqrt{\sigma_{\rm inst}^2 + \sigma_{\rm other}^{2}}$.

The data cubes representing each epoch are integrated over the disk's extant to create line profiles, $I_{\rm J}(V)$. Or,
\begin{equation}
    I_{\rm J}(V) = f_{\rm norm}\int_{}^{} F_{\rm J} \, G(V-V_{\rm p}) \,{\rm d} A \,.
    \label{eq:line_profile}
\end{equation}
$f_{\rm norm}$ is a normalization constant 
that also includes unit conversion between $F_{\rm J}$ and $G$. The normalization makes it so that the absolute value of the integrated line fluxes were never calculated explicitly. The SA signals corresponding to Eqn. \ref{eq:line_profile} are calculated by multiplying the flux densities with their projected distances along the PA slit, integrating and then dividing by the line profile. Or,
\begin{equation}
    {\rm SA}_{\rm J}(V) = \frac{\int_{}^{} F_{\rm J} \, r_{\rm p} \cos(\Omega) \, G( V-V_{\rm p}) \,{\rm d} A}{\int_{}^{} F_{\rm J} \, G(V-V_{\rm p}) \, {\rm d} A} \,.
    \label{eq:spectroastrometric_signal}
\end{equation}
$\Omega$ is the angle between the slit PA and the on-sky projected radii, $r_{\rm p}$. Eqn. \ref{eq:power_laws}-\ref{eq:spectroastrometric_signal} are performed for the relevant transitions (See Sec. \ref{sec:OBS}) where the individual line profiles and SA signals are averaged together for comparison with the data. In order words, we perform a fit to the averaged profiles presented in the bottom left panel of Fig. \ref{fig:ABAUR_all}.

The best-fit solution was determined from an overall reduced chi-squared, $\chi^{2}_{\rm red}$ accommodating all the fitted PAs. The chi-squared statistic value for an individual PA, $\chi^{2}_{i}$ where the subscript ``i'' denotes different PAs, was first calculated with
\begin{equation}
    \chi^{2}_{\rm i} = \sum_{\rm n=0}^{51} \frac{(\rm data_{n,i} - model_{n,i})^{2}}{\sigma_{\rm n,i}^2},
    \label{eq:chisquaredPA}
\end{equation}
where the summation (n) is over the velocity channels and $\sigma_{\rm n,i}$ is the standard deviation of the PA continuum. $\chi^{2}_{\rm red}$ is calculated by summing each $\chi^{2}_{i}$ and dividing by the total degrees-of-freedom, DoF. Or,
\begin{equation}
    \chi^{2}_{\rm red} = \sum_{\rm PA=0}^{3} \frac{\chi^{2}_{\rm i}}{\rm DoF},
    \label{eq:chisquaredreduced}
\end{equation}
where DoF = $N_{\rm pts}$ - $N_{\rm par}$ - 1, $N_{\rm pts}$ is the number of data points and N$_{\rm par}$ is the number of parameters fitted. We used $N_{\rm pts}$ = 204 because each PA has 51 unsmoothed data points each with a channel width of about 1~km~s$^{-1}$, and 4 PAs were fitted (51$\times$4=204). 

\begin{deluxetable*}{cccccl}[t]
\tablecaption{Best-fit parameters \label{tab:BESTFIT}}
\tablewidth{0pt}
\tablehead{\colhead{Fitted Transitions} & \colhead{Model Parameter} & \colhead{Best-Fit Value} & \colhead{Physical Parameter} & \colhead{Translated Value} }
\startdata
high-J & $\log_{10}(r_{\rm in})$  & 0.029$^{+0.059}_{-0.083}$ & $r_{\rm in}$  & 1.1$^{+0.2}_{-0.2}$~au          \\
       & $\log_{10}(r_{\rm out})$ & 1.77$^{+0.08}_{-0.07}$ & $r_{\rm out}$ & 59.3$^{+11.2}_{-8.7}$~au           \\
       & $\log_{10}(T_{\rm 0})$   & 3.50$^{+0.07}_{-0.05}$ & $T_{\rm 0}$   & 3140$^{+510}_{-370}$~K             \\
       & $T_\alpha$              & 0.11$^{+0.02}_{-0.02}$ & --            & ---                                \\
       & $\log_{10}(N_{\rm 0})$   & 18.1$^{+0.2}_{-0.20}$   & $N_{\rm 0}$   & 1.01$^{+0.74}_{-0.29}\times10^{18}$~cm$^{-2}$ \\
       & $N_\alpha$              & 1.93$^{+0.04}_{-0.04}$    & --            & ---                                \\
\hline
low-J & $\log_{10}(r_{\rm in})$   & 0.17$^{+0.05}_{-0.09}$ & $r_{\rm in}$  & 1.5$^{+0.2}_{-0.3}$~au \\
       & $\log_{10}(r_{\rm out})$ & 2.08$^{+0.12}_{-0.22}$    & $r_{\rm out}$ & 120$^{+38}_{-48}$~au      \\
       & $\log_{10}(r_{\rm s})$   & 1.81$^{+0.15}_{-0.08}$    & $r_{\rm s}$   & 64.6$^{+26.6}_{-10.9}$~au  \\
       & PA$_{\rm s}$            & 147$^{+3}_{-7}$~deg       & --            & ---                       \\
       & $\log_{10}(I_{\rm s})$         & 29.5$^{+0.7}_{-0.6}$    & $I_{\rm s}$         & 3.2$\times10^{29}$$^{+0.2}_{-0.1}$~erg~s$^{-1}$ \\
\enddata
\centering
\tablecomments{See Sec. \ref{sec:LINES_HI} for a description of the high-J model and Sec. \ref{sec:LINES_LO} for the low-J.}
\end{deluxetable*}

If we are to assume a chi-squared distribution, $\chi^{2}_{\rm dist}$, then the population mean is $\mu(\chi^{2}_{\rm dist}) = \rm DoF$ and the population standard deviation is $\sigma(\chi^{2}_{\rm dist}) = \rm \sqrt{2\times DoF}$. Using $\rm DoF=197$, we have $\mu(\chi^{2}_{\rm dis}) = \rm 197$ and $\sigma(\chi^{2}_{\rm dis}) = \rm$ 19.8, or about 10\% of $\mu(\chi^{2}_{\rm dis})$. The error bars of our fitted parameters are determined by finding the range of each parameter that falls within the 1$\sigma$ range of our chi-squared distribution.

Our approach for estimating the error bars does assume that the parameters are not co-dependent; in actuality, there are likely some co-dependencies. For example, the population density $N_{\rm j}$ from Eqn. \ref{eq:NJ} is degenerate with different combinations of surface number density $N$ and temperature $T_{\rm rot}$ (Eqn. \ref{eq:power_laws}). 
Consequently, the error bars are underestimated and should be taken with caution.
A detailed statistical study, such as deriving a Markov Chain Monte Carlo (MCMC) posterior distribution (see \cite{Sharma2017})
would be too computationally expensive with our current software and hardware.
%

The best-fit line profiles to the ${}^{12}$CO high-J transitions are shown in Fig. \ref{fig:bestfit_hiJ} (bottom) along with the corresponding SA signals (top). Table \ref{tab:BESTFIT} gives the best-fit values and error bars. Some parameters were fitted for efficiency using their log values (base 10), but the translated physical values are also given. The fitting routine was applied only to the line profiles and not the SA signals, which didn't have strong enough signals to warrant it. Regardless, the model SA signals appear to replicate the data in when discernible. Overall, the model sufficiently replicated the fitted line profiles for every PA where our fit gives $\chi^{2}_{\rm red}$ = 1.26.

The high-energy gas emission extends from $r_{\rm in}$ = 1.07~au to $r_{\rm out}$ = 59~au. These values may seem different from those presented in \cite{Jensen2024} ($r_{\rm in}$ = 1.0~au and $r_{\rm out}$ = 106~au), however, that study likely derived slightly different radii, namely $r_{\rm out}$, because they stacked both high- and low-J lines. We specifically chose to study the high-J lines to avoid substructure contamination present in the low-J lines (Sec. \ref{sec:high_vs_low}).

The best-fit parameters to the temperature power law from Eqn. \ref{eq:power_laws} are approximately $T_{0}$ = 3140~K and $T_{\alpha}$ = 0.11. When extrapolated to the radial bounds then $T(r_{\rm in})$ = 3140~K and $T(r_{\rm out})$ = 2030~K. Considering that the emission extends out to 59~au, it is curious that the disk surface retains a high enough temperature and alludes to it being heated by additional means. Further discussion is reserved for Sec. \ref{sec:DISS}.

The best-fit parameters to the surface density power law from Eqn. \ref{eq:power_laws} are approximately $N_{0}$ = 1.01e18 cm$^{-2}$ and $N_{\alpha}$ = 1.93. At the radial bounds this translates to $N(r_{\rm in})$ = 1.01$\times10^{18}$~cm$^{-2}$ and $N(r_{\rm out})$ = 4.32$\times10^{14}$~cm$^{-2}$. When looking at the model's calculations for the radial averages of the transitional optical depths (Eqn. \ref{eq:tau}), the disk's emission becomes optically thin ($\tau < 1.0$) for the high-J lines around 1.3~au (see also \cite{Miotello2014}). The optically thick-to-thin transition paired with the fitted temperature profile agrees with our observations in Fig. \ref{fig:rotational_temperature} (Sec. \ref{sec:RESULTS}).

\begin{figure*}[t]
    \centering
    \includegraphics[clip, trim = 14.5cm 6.0cm 16.5cm 11.0cm, width=0.99\textwidth]{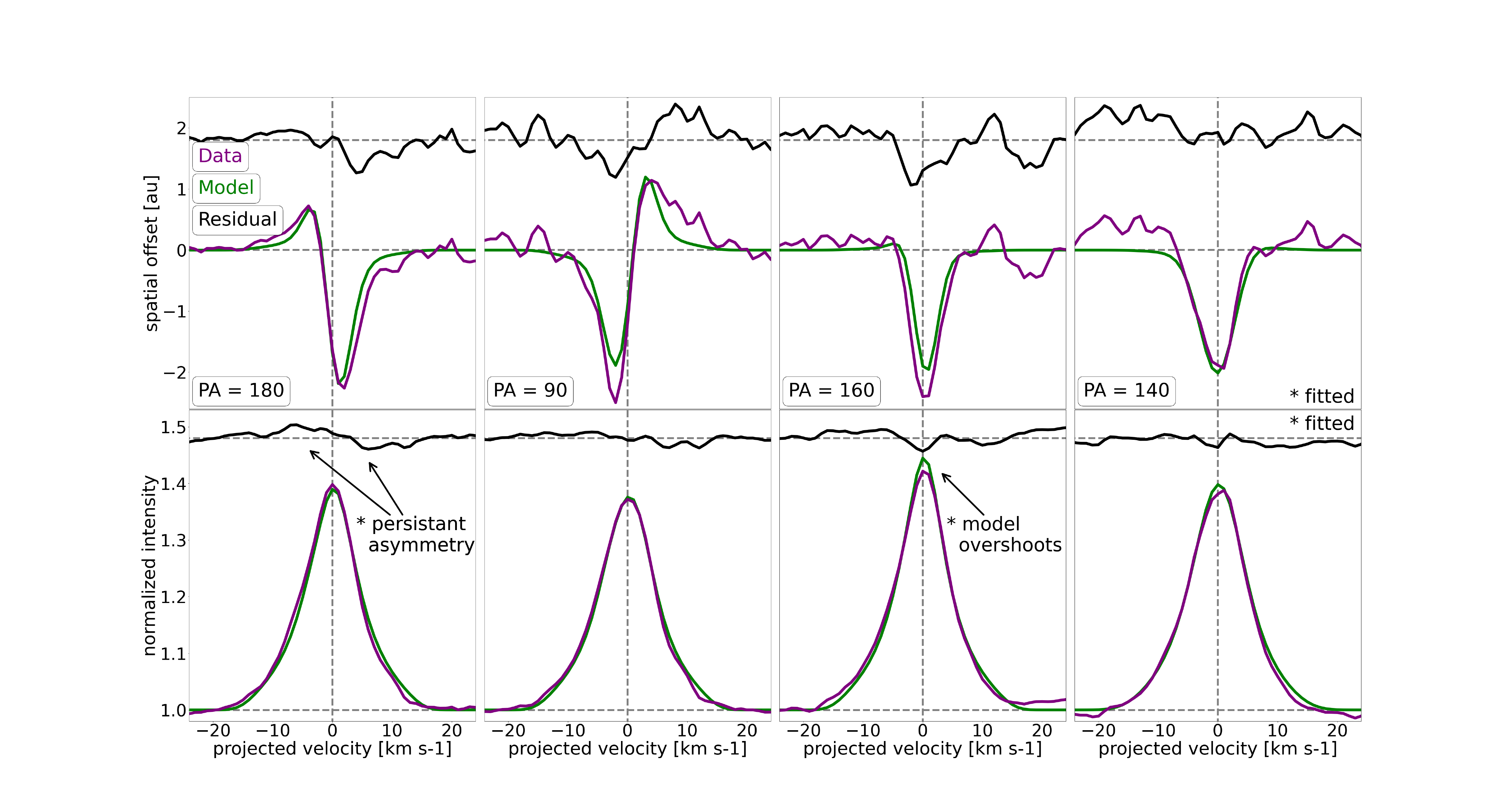}
    \caption{The low-J average transitions (purple) from Fig. \ref{fig:ABAUR_all} with the best-fit results (green). The analysis of these line profiles and spectroastrometric signals are described in Sec. \ref{sec:LINES_LO} and Sec. \ref{sec:LINES_LO}. By assuming the protoplanetary disk has an additional source of emission the flux and location of said source is constrained (Table \ref{tab:BESTFIT}).}
    \label{fig:bestfit_loJ}
\end{figure*}

The best-fit values for the surface temperature and density profiles ($T_{0}$/$T_{\alpha}$ and $N_{0}$/$N_{\alpha}$: Eqn. \ref{eq:power_laws}) will be held fixed in the subsequent analysis of the low-J lines (Sec. \ref{sec:LINES_LO}). Doing this assumes that high-J and low-J transitions originate from the same emitting layer but this simplification allows us to significantly reduce the dimensionality of our fit, which would not be feasible otherwise.

\subsubsection{Point Source Model (Low-J Fit)}
\label{sec:LINES_LO}
The SA signals of the low-J lines (top right; Fig. \ref{fig:ABAUR_all}) show a significant offset that suggests an off-centered, low temperature, and compact emission source located somewhere in the south-western portion of the protoplanetary disk (Sec. \ref{sec:high_vs_low}). Also, the line profile for PA160 seems to have captured more emission than the other PAs, possibly because this slit is more aligned in the source's direction. \cite{Jensen2024} made similar comments but performed no specific analysis. In this section, to test the source's compactness, we parameterize it as a point source that is inhabiting the disk.

Similar to Sec. \ref{sec:LINES_HI}, the emission of the protoplanetary disk is calculated using Eqns. \ref{eq:power_laws}-\ref{eq:transitional_flux}, except that the temperature and surface density parameters in Eqn. \ref{eq:power_laws} are fixed to the best-fit values presented in Table \ref{tab:BESTFIT}. Consequently, the only means by which the protoplanetary disk can change its intensity is by altering the inner and outer radii, $r_{\rm in}$/$r_{\rm out}$. 

A point source with intensity $I_{\rm s}$ is inserted onto the protoplanetary disk at radius $r_{\rm s}$ and PA$_{\rm s}$, adding an extra term to Eqn. \ref{eq:transitional_flux}:
\begin{equation}
    F_{\rm J} = (1-e^{-\tau_{\rm J}})\,B_{\rm J}(\tilde{\nu}_{\rm J},T_{\rm rot}) + I_{\rm{s}}\delta(r_{\rm s},{\rm PA}_{\rm s}) \, ,
\end{equation}
where $\delta$ is the Dirac delta function. Numerically, the point source is superimposed via bilinear interpolation in our model grid, which inherently assigns an area to the source --- 1.25~au$^{2}$ in our case --- but this size is irrelevant since the seeing is several tens of times larger (Table \ref{tab:epochs}) . Also, the point source is assigned the corresponding orbital velocity at $r=r_{\rm s}$ and, for simplicity, we assume $I_{\rm s}$ is equal in all transitions.

As was done for Sec. \ref{sec:LINES_HI}, the model data cube is spatially convolved and applied with slit masks in accordance to the observing conditions (Table \ref{tab:epochs}). The constructed line profiles and SA signals (Eqn. \ref{eq:line_profile} and Eqn. \ref{eq:spectroastrometric_signal}) are spectrally convolved, normalized to the data, and fit to the averaged data presented in the right panels of Fig. \ref{fig:ABAUR_all}. Unlike in Sec. \ref{sec:LINES_HI}, which focused solely on the line profiles, here we minimize the equally weighted sum of $\chi^{2}_{\rm red}$'s of both the line profiles and SA signals. The best-fit line profiles and SA signals are presented in Fig. \ref{fig:bestfit_loJ} with the best-fit parameters given in Table \ref{tab:BESTFIT}. 

The best-fit line profiles and SA signals to the ${}^{12}$CO low-J transitions (Fig. \ref{fig:bestfit_loJ}) adequately replicated the data and, using Eqn. \ref{eq:chisquaredreduced}, has a $\chi^{2}_{\rm red}$ value of 0.97 for the line profiles. The best-fit parameters are presented in Table \ref{tab:BESTFIT}, and the error bars were estimated similarly to Sec. \ref{sec:LINES_HI}. However, the errors for $r_{\rm s}$ and PA$_{\rm s}$ are based on the reduced chi-squared of the SA signals --- which is 1.42. These parameters are coordinates whose immediate effects are seen in the SA signals.

Generally, the fitting routine decently replicated the line profiles and SA signals across all four PAs (Fig. \ref{fig:bestfit_loJ}). However, even though our best-fit profiles were nearly symmetric around the line cores, a consistent high-velocity asymmetry remained unaddressed. We see that the blue-shifted wings ($\sim-15$~km~s$^{-1}$) overshoots the model while the red-shifted wings ($\sim+15$~km~s$^{-1}$) undershoots it. As discussed in Sec. \ref{sec:high_vs_low}, this behavior is not seen in the high-J lines, and its origin is unclear. Further discussion is reserved for Sec. \ref{sec:DISS}.

The best-fit of the protoplanetary disk's inner and outer radii (Table \ref{tab:BESTFIT}) are $r_{\rm in}$ = 1.49~au and $r_{\rm out}$ = 120~au. These are closer to the values derived by \cite{Jensen2024}. When compared to the radial bounds determined in Sec. \ref{sec:LINES_HI}, they are near the 1$\sigma$ uncertainty but some discrepancy is expected as those were derived from different transitions which get excited with differing conditions. The CO emission is less extended than the dust ring \citep{Tang2012}, but its 120~au extent may be consistent since CO gas around AB Aur has been detected at many hundreds of au \citep{Pietu2005, Corder2005} and the surface of protoplanetary disks around Herbig stars can remain warm at large distances \citep{Thi2013}.

\begin{figure}[t]
    \centering
    \includegraphics[clip, trim = 1.0cm 0.8cm 3.2cm 3.5cm, width=0.99\linewidth]{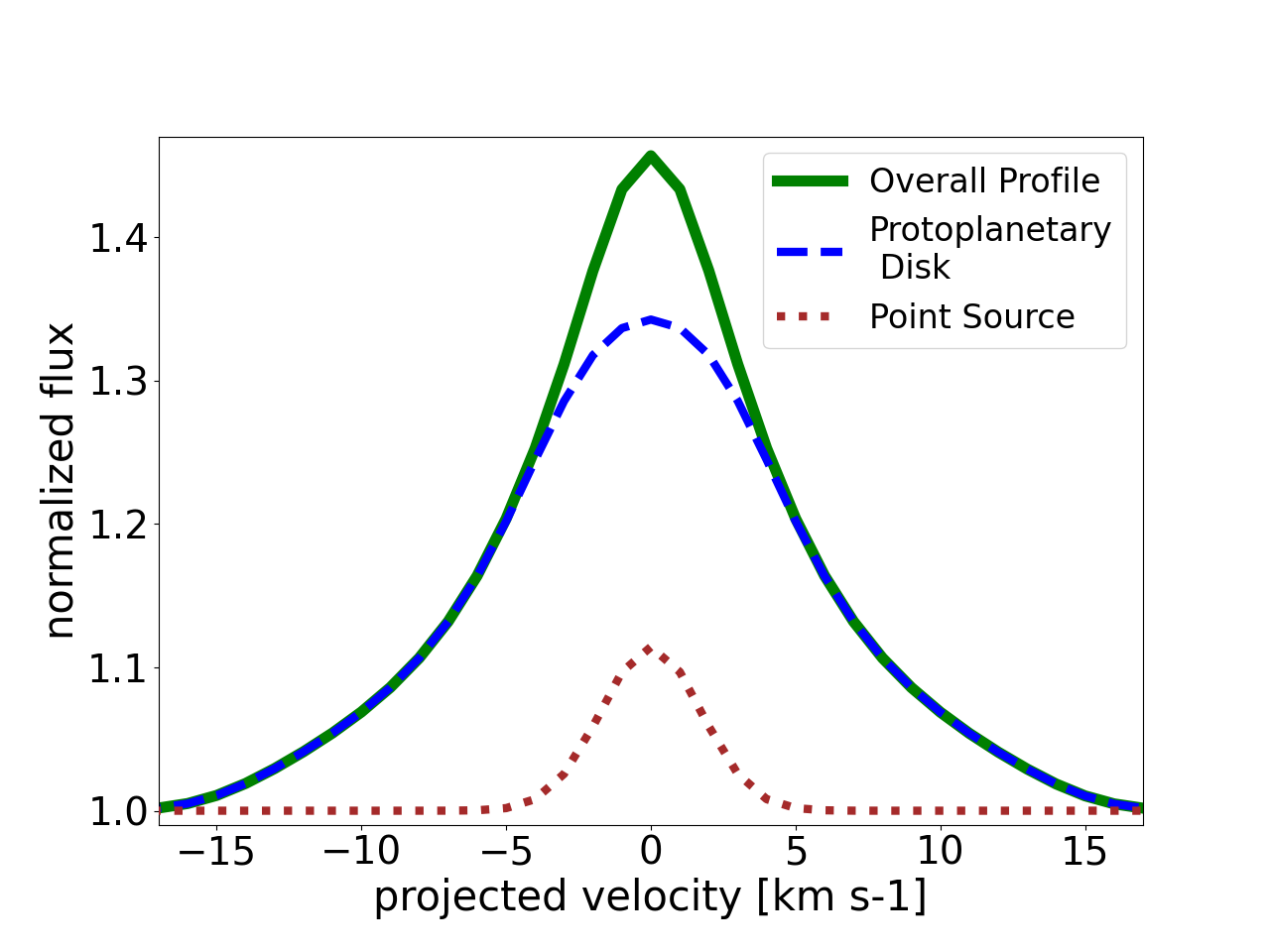}
    \caption{The component contributions (Sec. \ref{sec:LINES_LO}) of the protoplanetary disk (blue) and the modeled point source (brown). This image is constructed from the best-fit results of PA160 in Fig. \ref{fig:bestfit_loJ}. The equivalent width of the point source is roughly 10\% of the protoplanetary disk's and this information is used to calculate the point source's temperature in Sec. \ref{sec:LINES_HI}.}
    \label{fig:components}
\end{figure}

The point source is located around $r_{\rm s}$ = 65~au and PA$_{\rm s}$ = 147\degr which gives it a projected velocity of around +0.2~km~s$^{-1}$. This small velocity is enough to make the overall line profile appear symmetric. The coordinates are fully captured by both the PA160 and PA140 slits (see Fig. \ref{fig:Observations}), but PA160 has almost double the spectral resolution. Arguably, PA160 contains the most information about the source. As such, Fig. \ref{fig:components} and Fig. \ref{fig:substructures} (discussed later) are constructed from the model's PA160.

The derived intensity of the source was determined to be $I_{\rm s}$ $\approx$ 3.2$\times10^{29}$~erg~s$^{-1}$ per low-J line. Fig. \ref{fig:components} is the best-fit line profile of PA160 from Fig. \ref{fig:bestfit_loJ} along with the component profiles. The equivalent width of the point source's sub-profile is $\approx$10\% of the overall profile.

The source's non-detection in the high-J transitions (Sec. \ref{sec:high_vs_low}) constrains its temperature. Assuming optically thin emission, the source's flux, $F_{\rm s}$, can be ratioed between its high-J and low-J variants ($F_{\rm s,hi}$/$F_{\rm s,hi}$) and approximates to a ratio of optical depths (Eqn. \ref{eq:tau}). Or,
\begin{equation}
    \label{eq:tau_ratio}
    \frac{F_{\rm s,hi}}{F_{\rm s,lo}} \approx \frac{\tau_{\rm hi}}{\tau_{\rm lo}} =
     \frac{N_{\rm hi}}{N_{\rm lo}}\frac{g_{\rm hi} g^{\prime}_{\rm lo}}{g^{\prime}_{\rm hi} g_{\rm lo}} \frac{A_{\rm hi}}{A_{\rm lo}}\frac{\tilde{\nu}_{\rm lo}^3}{\tilde{\nu}_{\rm hi}^3} .\
\end{equation}
The subscripts 'hi' and 'lo' refer to the chosen high-J and low-J transitions used for the calculations --- which are P(27) and P(3). The surface density ratio, $N_{\rm hi}$/$N_{\rm lo}$, is the ratio of Eqn. \ref{eq:NJ};
\begin{equation}
    \frac{N_{\rm hi}}{N_{\rm lo}} = \frac{g_{\rm hi}}{g_{\rm lo}} \exp\left(\frac{-(E_{\rm hi} - E_{\rm lo})}{ k_{\rm B} T_{s}}\right) ,\
\end{equation}
where $T_{\rm s}$ is the source's temperature. Based on our equivalent width measurements and errors (Table \ref{tab:EQW}), we estimate that the source should contribute no more than 4\% than what was measured in the high-J transitions; or $F_{\rm s,hi}$ / $F_{\rm hi} \leq$ 0.04. It was determined above that the point source contributes 10\% to the low-J transitions; or $F_{\rm s,lo}$ / $F_{\rm lo}$ = 0.10. Using these estimates, the ratio of the source's fluxes (Eqn. \ref{eq:tau_ratio}) should satisfy
\begin{equation}
    \label{eq:flux_ratio}
    \frac{ F_{\rm s,hi}}{ F_{\rm s,lo}} \lesssim \frac{ 0.04 F_{\rm hi}}{ 0.10F_{\rm lo}} = 0.24 \,.
\end{equation}
Solving for $T_{\rm s}$ after equating Eqn. \ref{eq:tau_ratio} and Eqn. \ref{eq:flux_ratio} gives $T_{\rm s}\lesssim$550~K. 
The same line ratio is not attainable if we instead assume the lines are all optically thick, which strongly suggests the source is optically thin in its high-J transitions.

The results of this section treated the off-centered emission as if it originates from a point source. In Sec. \ref{sec:LINES_LOCPD}, an alternative model is tested to see how representative a point source is and discussion of the substructure's nature is reserved for Sec. \ref{sec:DISS}.

\subsubsection{Circumplanetary Disk Model (Low-J Fit)}
\label{sec:LINES_LOCPD}
In this section, we check if the off-centered source may resemble a CPD rather than a point source (Sec. \ref{sec:LINES_LO}) by applying another fitting routine to the low-J transitions (right: Fig. \ref{fig:ABAUR_all}). If discernible, the two main distinguishing features a CPD will have are 1) its definable area and 2) its characteristic velocity spread. However, if the planet's mass is below a specific value, it may be indistinguishable from a point source in our analysis.

Assuming the planet's orbit is circular and coplanar, the CPD is created using the same algorithm for the general protoplanetary disk model (Sec. \ref{sec:LINES_HI}) except the stellar mass in Eqn. \ref{eq:projected_velocity} is replaced with the planetary mass parameter, $M_{\rm P}$. The planet's mass also defines the CPD's area and rotation rates. The CPD's radius, $R_{\rm CPD}$, is set to be one-half of the Hill Sphere radius, $R_{\rm H}$ (see \cite{Machida08}). Or,
\begin{equation}
    R_{\rm CPD} = 0.5 R_{\rm H} = 0.5 r_{\rm s} \left(\frac{M_{\rm P}}{3M_{*}}\right)^{\frac{1}{3}} \,.
    \label{eq:RH}
\end{equation}
This description for the CPD radius may be more representative of massive planets but it may obey a different scaling law for lower mass planets (see \cite{Fung2019}). Using a different prescription than that in Eqn. \ref{eq:RH} may result in a different planet mass, but would not change the area of the emitting source.

\begin{figure}[b]
    \centering
    \includegraphics[clip, trim = 0.2cm 0.1cm 2.6cm 2.9cm, width=0.99\linewidth]{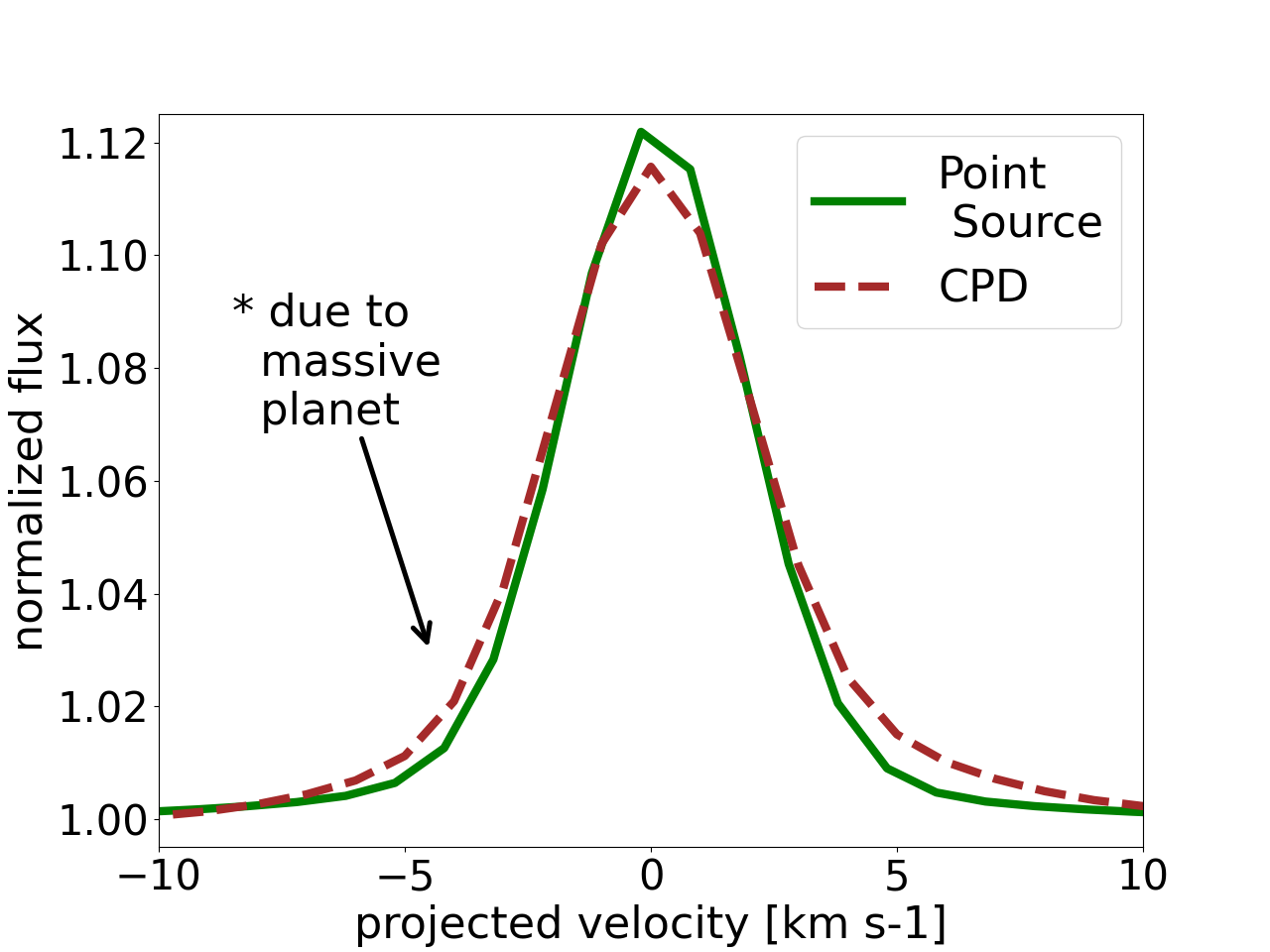}
    \caption{To fit the low-J transitions from Fig. \ref{fig:ABAUR_all} two models are constructed to determine if the secondary components are differentiable. One is of a point source (Sec. \ref{sec:LINES_LO}) and the other is of a circumplanetary disk (Sec. \ref{sec:LINES_LOCPD}). Though the differences are too small to be discernible in our analysis we do see that, given a massive enough planet, the circumplanetary disk's signature ends up being wider and smaller amplitude.}
    \label{fig:substructures}
\end{figure}

Because the CPD is far smaller than the seeing conditions (Tab. \ref{tab:epochs}), it is not meaningful to prescribe a brightness profile to the CPD. Instead, we assume a spatially constant flux, $\mathfrak{F}_{\rm CPD}$. This approach allows the CPD's line profile, $I_{\rm CPD}$, to be described by:
\begin{equation}
    I_{\rm{CPD}}(V) = \pi R_{\rm CPD}^{2} \mathfrak{F}_{\rm{CPD}} G( V-V_{\rm p}) \,.
\end{equation}
Utilizing this approach keeps the number of fitted parameters low and we do not expect our results to change if the CPD's flux was made more complicated.

Like the point source in Sec. \ref{sec:LINES_LO}, the CPD is superimposed on the protoplanetary disk, which is unchanged from Sec. \ref{sec:LINES_LO}, with the coordinates of $r_{\rm s}$ and PA$_{s}$. The master disk image is then convolved with seeing and masked with the appropriate observing slits (Table \ref{tab:epochs}). The resulting line profiles and SA signals are also spectrally convolved and normalized.

The CPD model optimized its 6 parameters --- $r_{\rm in}$, $r_{\rm out}$, $\mathfrak{F}_{\rm CPD}$, $r_{\rm s}$, PA$_{s}$, $M_{\rm s}$ --- according to the same conditions as the point source model with its 5 parameters. The reduced chi-squared (Eqn. \ref{eq:chisquaredreduced}) of the best-fit CPD model is quantitatively similar to those of the point source model, meaning that no significant improvement in the results were obtained.

Even though the two models ended up being similar, certain features appear in the CPD model that are worth pointing out. The component profiles of the CPD and point source are presented in Fig. \ref{fig:substructures}. The CPD's profile has a smaller amplitude and a wider spread than the point source's. But this feature only becomes apparent with a large planet mass. If the planet's mass is small, then the CPD's profile becomes the same as the point source's; whose width is characterized by $\sigma_{\rm Gauss}$. Though it does look like the differences may be important, we would like to remind the reader that the profiles in Fig. \ref{fig:substructures} are small contributions to the overall profile. 

The best-fit planet mass parameter is $M_{\rm P}$ $\approx$ 20~$M_{\rm J}$ which, using Eqn. \ref{eq:RH}, corresponds to a CPD radius of almost 5~au. Given that our CPD fit is statistically indistinguishable from our point source fit, 5~au can be considered an upper limit to the source's radius. A CPD of this size may also be too small to be captured using imaging techniques. \cite{Rab2019} suggested that a CPD may need a radius of roughly 10~au to be detectable by ALMA after 6 hours of observation on a target system that is roughly 150~pc away. 


Overall, although slight differences exist between our CPD model and the point model (Fig. \ref{fig:substructures}), they remain very similar and did not result in meaningfully different fits. Therefore, we cannot conclusively determine whether this CO source is a spinning, extended object, like a CPD. On the other hand, the possibility of it being a CPD is not ruled out. In the next section, we run a hydrodynamic simulation assuming the source is a planet to present possible protoplanetary disk features that may arise.

\subsection{Hydrodynamic Simulation}
\label{sec:HYDRO}
This section presents a hypothetical thought experiment that can potentially relate our detected source, assumed to be a planet, with AB Aur b, which will be treated as a disk feature for this section. A hydrodynamic simulation is performed using \texttt{PEnGUIn} \citep{Fung2015} in a manner similar to what was presented in \cite{Kozdon2023}. The simulation is defined on a two-dimensional, radial-azimuthal grid, with a resolution of $\sim9$ cells per scale height. We assume a disk aspect ratio of 7.5\% and a Sunyaev-Shakura alpha viscosity parameter of $\alpha=10^{-4}$. These values are typical for protoplanetary disks at tens of au. The planet is placed on a fixed, circular, and coplanar orbit.

AB Aur's high-J line profiles appear symmetric (Fig. \ref{fig:ABAUR_all}; Sec. \ref{sec:high_vs_low}), which informs us that the overall disk eccentricity is relatively low. As such, a planet mass must be chosen so that the disk eccentricity is not excited to a high level. A suggested companion-driven, disk eccentricity excitation planet-to-star mass ratio  threshold is about 0.003 \citep{Kley2006, Kanagawa2015, Tanaka2022}. Therefore, with a stellar mass of 2.4~$M_\odot$ \citep{Rodriguez2014}, we choose a planet mass of 5$M_{\rm J}$ to be below that limit.

\begin{figure}[t]
    \centering
    \includegraphics[clip, trim = 0.9cm 6.1cm 9.82cm 5.7cm, width=0.99\linewidth]{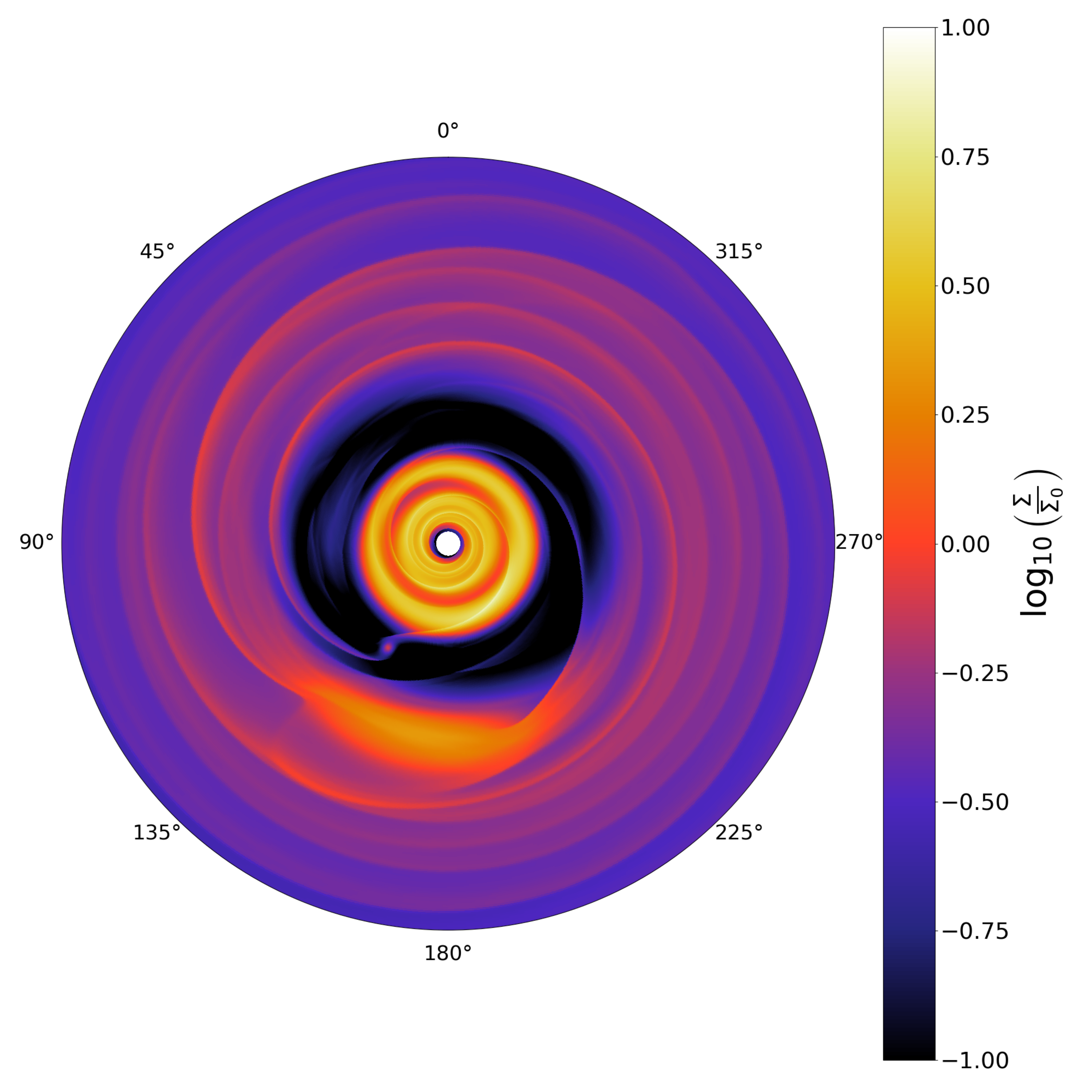}
    \caption{Plotted disk surface density of a hydrodynamic simulation (Sec. \ref{sec:HYDRO}) with a $5 M_J$ companion on a circular orbit. The disk has a $7.5\%$ disk-aspect ratio and a $10^{-4}$ $\alpha$-viscosity. This snapshot is of the $250th$ orbit. Color contour of the plot is normalized to the initial disk density at the planet's semi-major axis to be in code units of $1$, where the warmer the color the higher the disk density. In the snapshot the planet was able to carve a gap and induce a vortex to develop.}
    \label{fig:hydro_simulation}
\end{figure}

The simulation was run for hundreds of orbits, and the snapshot of the 250th orbit is presented in Fig. \ref{fig:hydro_simulation}. The warmer colors in the color gradient correspond to higher disk densities. Also, the image is rotated to be reminiscent of the CHARIS image of AB Aur \citep{Currie2022} seen in Fig. \ref{fig:Observations}.

In Fig. \ref{fig:hydro_simulation}, the planet carved a circular gap centered on its orbit (see \cite{Dong2017}). A vortex, generated by the Rossby Wave instability (see \cite{Lovelace2014}), is present just beyond the gap due south of the star, roughly where AB Aur b would be located. However, the planet and the vortex do not maintain this configuration for long as they both orbit the star with their respective Keplerian velocities, meaning their relative positions will change over time.

Though Fig. \ref{fig:hydro_simulation} could serve as a possible explanation between our detected source and AB Aur b, we stress that this is simply one of many possibilities. The two features may not even be related at all. Whether the newly detect source is of planetary origin remains an open question. 

\section{Discussion}
\label{sec:DISS}
When comparing the line profiles presented in Fig. \ref{fig:ABAUR_all}, it is quite peculiar that the high-velocity ($\pm$15~km~s$^{-1}$) asymmetry is captured only in the low-J transitions (Sec. \ref{sec:high_vs_low}). Typically, asymmetries can be indicative of substructures like circumplanetary disks \citep{Brittain2019}, eccentric annuli \citep{Kozdon2023, Liskowsky2012}, disk winds \citep{Pontoppidan2011} or vortices. However, the signatures of the mentioned substructures should be present in the majority of the excited transitions. The observed asymmetry being present only in the low-J transitions indicates some low-temperature inner disk features. What this could be is not clear to us and is left unresolved in this study.

In Sec. \ref{sec:LINES_HI}, the protoplanetary disk's surface temperatures from Eqn. \ref{eq:power_laws} was determined to have a high normalized temperature of $T_{0}$ $\approx$ 3100~K and a low power law of $T_{\alpha}$ $\approx$ 0.1, extending out to $\sim60$~au. The low-J emission extends even further to $\sim120$~au (Table \ref{tab:BESTFIT}). The sources of heat that may be maintaining this emission layer may include the interactions of polycyclic aromatic hydrocarbons \citep{Kamp2004}, which have been resolved in this system \citep{Yoffe2023}, cosmic ray interactions \citep{Glassgold2012}, and/or turbulent heating \citep{Kurbatov2024}. The disk's elevated temperature may also be related to its ability to cool. We determined that the surface density in the outer disk (beyond $\sim100$~au) is about $10^{14}$~cm$^{-2}$ and is likely optically thin (Sec. \ref{sec:LINES_HI}). This is consistent with previous observations that have determined little gas presence at similar distances \citep{Oppenheimer2008}. 
This low surface density and the fact that dust settling can deplete dust from the CO layer we observe, might lead to a weakened gas-dust thermal coupling and allow the gas to heat up.
The mechanism that creates the thermophysical environment around AB Aur deserves further investigations.

A hot protoplanetary disk may have implications for gravitational instability \citep{Su2025}. From studying AB Aur's outer disk ($r\ge$100~au) \cite{Speedie2024} found that the kinematic signature can be partially recreated by a model that simulates gravitational instability (see also \cite{Speedie2025}), and found that the disk may be as massive as a third of the star's mass. To prevent such a massive disk from fragmenting due to gravitational instability, the disk must be hotter than usual. Our observations are in line with this idea.

A compact source that inhabits the protoplanetary disk was modeled to address the variable emission between PA's observed in the low-J transitions (right: Fig. \ref{fig:ABAUR_all}). The compact nature was required because of the drastic offset captured in the SA signals between PAs. The observed behavior was explained by placing an orbiting object so that its projected velocity is near 0, which, in our case, was $r_{\rm s}$ = 65~au and PA$_{\rm s}$ = 147\degr (Sec. \ref{sec:LINES_LO}). One possibility is that the source is not on a circular orbit. This would open up a larger range of parameter space for its spatial location.

The low-J line profiles and SA signals (Fig. \ref{fig:bestfit_loJ}) were fitted with a model that was made up of a protoplanetary disk and an additional point source (Sec. \ref{sec:LINES_LO}). The point source was placed in the southeastern portion of the disk at $r_{\rm s}$ = 65~au and PA$_{\rm s}$ = 147$\degr$. Other point sources have been reported by \cite{Boccaletti2020}; one has coordinates of roughly $r = $25~au and PA$ = 204\degr$, while the other one is at around $r = 106$~au and PA$ = 8\degr$. We report that we did not detect these sources even though our slit arrangements (Fig \ref{fig:Observations}) should have captured them. Other studies like \cite{Speedie2025} and \cite{Bowler2025} have also failed to detect these sources. 

\subsection{It is not AB Aur b}
\label{sec:notABAurb}
\begin{figure}[t]
    \centering
    \includegraphics[clip, trim = 0.1cm 0.1cm 0.01cm 0.2cm, width=0.99\linewidth]{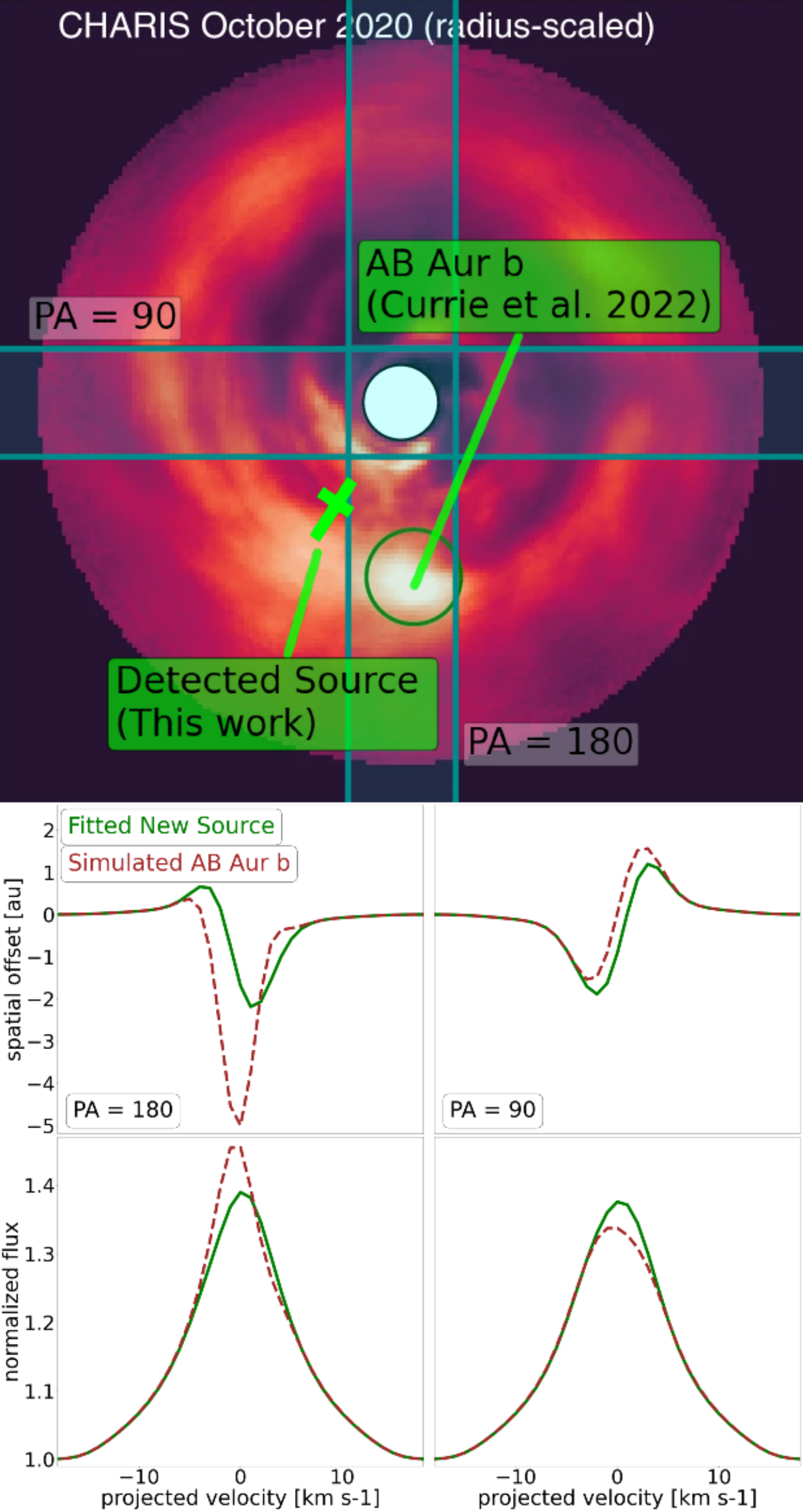}
    \caption{The CHARIS image from \cite{Currie2022} with AB Aur b and the detected source pointed out (top). The cross represents the new source's error. PA180 and PA90 are shown with the corresponding line profiles and spectroastrometric signals of the two sources plotted (bottom). The profiles/signals of the new source are pull from Fig. \ref{fig:bestfit_loJ} whereas those AB Aur b are simulated from our model. If detected, AB Aur b would have resulted in a larger offset in PA180. Also, all the line profiles would have developed a low-velocity asymmetry.}
    \label{fig:ABAur_sources}
\end{figure}

AB Aur b is a planetary candidate estimated to have a semi-major axis of roughly $r = 94$~au and a position angle of PA$\approx$180. With these coordinates, AB Aur b was theoretically captured in our observation (see Fig. \ref{fig:Observations}). Our results from Sec. \ref{sec:LINES_LO} localizes a point source at $r_{\rm s}$= 65~au and PA$_{\rm s}$ = 147$\degr$. Even when the error bars are considered, our source does not overlap with AB Aur b. See Fig. \ref{fig:ABAur_sources} for a comparison of AB Aur b and our newly detected source. At the top is a schematic of the CHARIS image from \cite{Currie2022} with lines pointing at the discussed sources. The orthogonal slits of PA90 and PA180 are overlayed to highlight the relative coverage. The bottom portion of Fig. \ref{fig:ABAur_sources} are the line profiles and SA signals that would correspond to the sources from the given PAs. The profiles and signals of our detected source comes from the results of Sec. \ref{sec:LINES_LO} (see Fig. \ref{fig:bestfit_loJ}). For AB Aur b, we simulated its resultant profiles and signals by running our model with the same values for the parameters $r_{\rm in}$, $r_{\rm out}$ and $I_{\rm s}$ (Table \ref{tab:BESTFIT}) but changed $r_{\rm s}$ and $PA_{\rm s}$ to those of AB Aur b. 

Fig. \ref{fig:ABAur_sources} highlights the key differences in the kinematic signatures between our detected source and AB Aur b. Note the significantly larger offset in PA180's SA signal. Also, across multiple PAs AB Aur b's orbital velocity caused the line profiles to become more asymmetric around the low projected velocities ($\leq\pm$10~km~s$^{-1}$). This asymmetry differs from the high-velocity asymmetry discussed in Sec. \ref{sec:high_vs_low}, which our model cannot address. Because of the differences in the line profiles and SA signals that AB Aur b would have introduced, we do not think it could have satisfactorily explained the collected data. As such, we conclude that our detected source is different from AB Aur b.

If AB Aur b emits any significant $^{12}$CO ro-vibrational lines, the protoplanetary disk should not be able to shield it, given that our fit determines $^{12}$CO is optically thin at this distance. The fact that AB Aur b is absent in our observations may mean that it is simply not emitting, which could be either due to it 1) being too cold or 2) being CO depleted. On the other hand, if it is emitting, the emission could be shielded by dust in the disk (but not by gas since CO should be optically thin (Sec. \ref{sec:LINES_HI}) at this location), which would imply the gas structure around AB Aur b is well embedded inside the dust layer. Further investigation is required to understand why AB Aur b is not seen.

From our inspection of the literature, we have not found an obvious candidate for what we have detected. Multiple spiral arms have been reported \citep{Boccaletti2020, Tang2017, Lin2006}, but none seemingly match. The detected source also does not seem close enough to the potential dust trap in the eastern portion of the dust ring \citep{Fuente2017, Tang2012}. We also believe our source is not a disk wind as that is a substructure that is more extended than our results suggest. There is the possibility that we detected a wholly new substructure, and because it is consistent with being a point source (Sec. \ref{sec:LINES_LO} and \ref{sec:LINES_LOCPD}), it may be of planetary origin.

\section{Summary and Conclusions}
\label{sec:CONC}
AB Aur was observed for 3 epochs with NASA IRTF in January 2024 with multiple position angles (Fig. \ref{fig:Observations}; Table \ref{tab:epochs}; Sec. \ref{sec:OBS}) to capture varying emission from a non-axisymmetric source. In Sec. \ref{sec:RESULTS}, the stacked emission line profiles and spectroastrometric signals of the M-band ${}^{12}$CO ro-vibrational high-J and low-J transitions (Fig. \ref{fig:ABAUR_all}; Sec. \ref{sec:high_vs_low}) as well as the hydrogen Pf$\beta$ transition (Fig. \ref{fig:hydrogen}; Sec. \ref{sec:hydrogen_transition}) are presented. The differing characteristics of the line profiles and spectroastrometric signals between the high-J and low-J ${}^{12}$CO transitions are compared (Sec. \ref{sec:high_vs_low}) and fitted with disk models (Sec. \ref{sec:LINES_FIT}). Where the thermophysical environment of the emitting layer is constrained (Sec. \ref{sec:LINES_HI}) and a substructure is potentially characterized (Sec. \ref{sec:LINES_LO} and Sec. \ref{sec:LINES_LOCPD}). A hydrodynamic simulation is run to study the potential outcomes of companion interactions with the disk (Sec. \ref{sec:HYDRO}). Lastly, our results our discussed in Sec. \ref{sec:DISS} where comparisons were made with the protoplanet candidate AB Aur b (Sec. \ref{sec:notABAurb}). Our main conclusions are highlighted below:

\begin{itemize}
    \item The stacked emission line profiles of the ${}^{12}$CO low-J transitions vary with PA where, in contrast, those for the high-J transitions did not (Fig. \ref{fig:ABAUR_all}). The low-J spectroastrometric signals also displayed considerable offsets which would arise from an off-center source of emission. These observations allude to a low-temperature substructure being present in the southwestern portion of the protoplanetary disk.

    \item The fitting routine of the low-J transitions (Fig. \ref{fig:bestfit_loJ} places the substructure at around $r_{\rm s}$ = 65~au and PA = 147$\degr$ (Sec. \ref{sec:LINES_LO}). While its origin is unknown, we are confident that it is not AB Aur b; in fact, emission from AB Aur b was not detected (Sec. \ref{sec:notABAurb}).

    \item The source is likely optically thin in $^{12}$CO high-J transitions, with a maximum temperature of $\sim550$~K (Sec. \ref{sec:LINES_LO}) and a maximum radius of $\sim5$~au (Sec. \ref{sec:LINES_LOCPD}).
    
    \item The strength of hydrogen Pf$\beta$ transition remained consistent over a two-decade cadence (Sec. \ref{sec:hydrogen_transition}) as well as over our one week observation run (Fig. \ref{fig:hydrogen}). However, this behaviour is at odds with previous studies that reported large changes over the course of one year. The calibrated stellar accretion rate is $\dot{ M} \approx 2.0\times10^{-7} M_{\sun}$ yr$^{-1}$ (Sec. \ref{sec:hydrogen_transition}).

    \item A persistent asymmetry was captured in the low-J line profiles (Sec. \ref{sec:high_vs_low}) that remained unaddressed in our analysis.
\end{itemize}

The analysis presented in this study cannot conclusively determine the nature of the detected source. One way to characterize it is by how it changes the line profiles over the years. Suppose PAs similar to those in Table \ref{tab:epochs} are observed later and the line profiles (Fig. \ref{fig:ABAUR_all}) evolve according to an orbiting body then that can indicate that the detected source is a planet or some other orbiting feature. 

Our analysis of the ${}^{12}$CO emissions (Sec. \ref{sec:LINES_FIT}) utilized a simple two-dimensional slab model that does not adequately account for three-dimensional effects such as disk flaring \citep{diFolco2009, Brittain2003} or vertical structure variations \citep{Law2022}. Future analysis may incorporate a more rigorous disk geometry to capture more complexities (see \citep{Kozdon2023}). Also, the fact that we studied the average profiles instead of the spectra may have lead to the loss of some information. Future analysis can do that as well as treat the source's emission as non-constant between transitions.

\section{Acknowledgments}
We thank an anonymous referee for suggestions that improved this manuscript. This work includes data gathered at the Infrared Telescope Facility, which is operated by the University of Hawaii under contract with the National Aeronautics and Space Administration (80HQTR19D0030). This work was supported primarily by the South Carolina Space Grant Consortium Graduate Research Fellowship under contract with the National Aeronautics and Space Administration (80NSSC20M0054). This work was supported in part by the South Carolina Space Grant Consortium under contract with the National Aeronautics and Space Administration (80NSSC22M0064). Y.H. was supported by the Jet Propulsion Laboratory, California Institute of Technology, under contract with the National Aeronautics and Space Administration (80NM0018D0004).

\bibliography{manuscript.bib}
\bibliographystyle{aasjournal}

\newpage
\appendix
\section{The Multigrid Setup}
\label{sec:Appendix}
\begin{figure}[h]
    \centering
    \includegraphics[clip, trim = 5.5cm 1.5cm 6.0cm 2.5cm, width=0.49\linewidth]{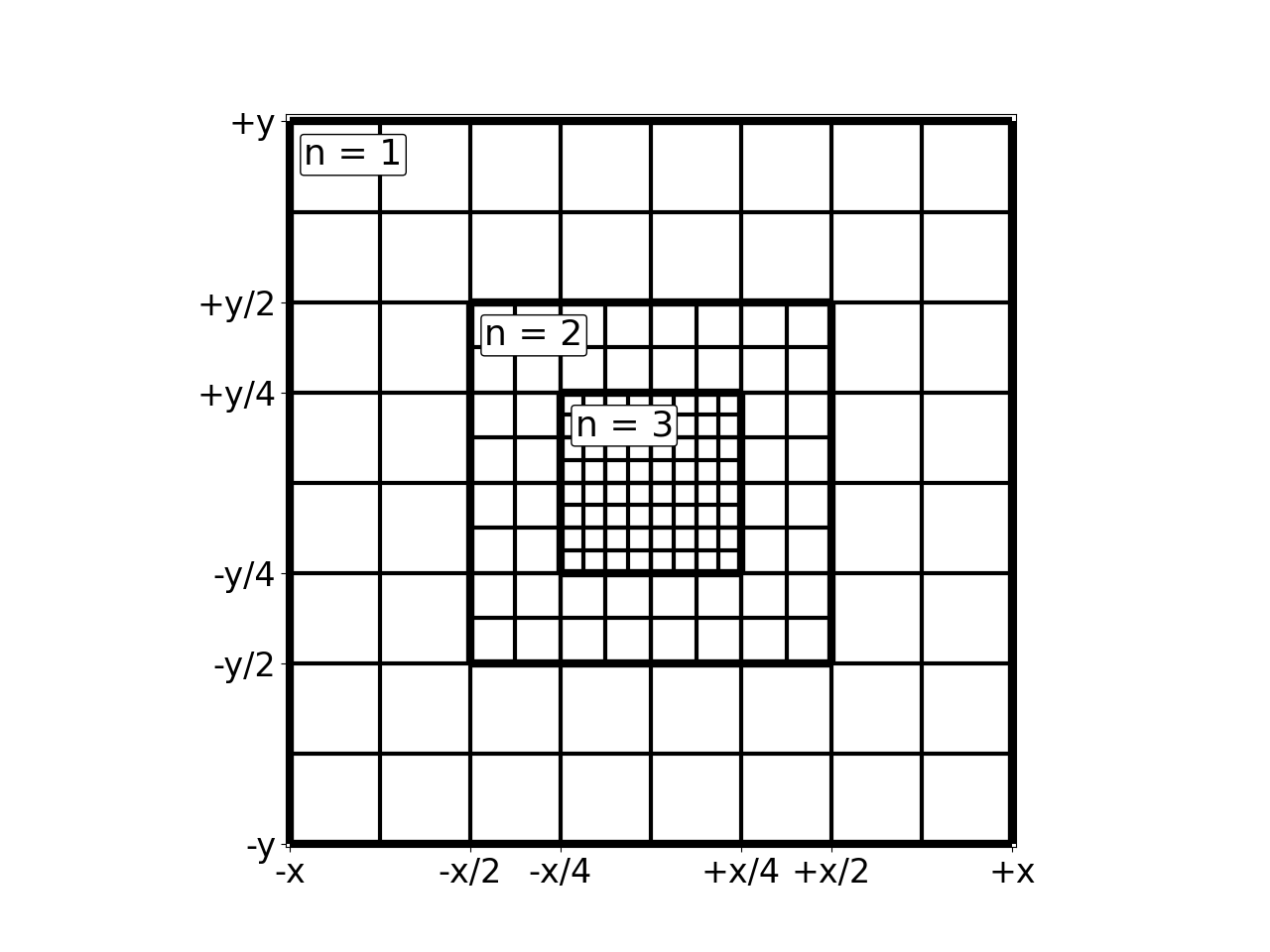}
    \caption{An example of a lower resolution multi-grid that can be employed for this study (Sec. \ref{sec:LINES_FIT}). The multi-grid has increased resolution towards the center which is where the higher velocities of a protoplanetary disk are. The above example has 3 concentric 8x8 grids whereas the ones used for modeling has 15 concentric 192x192 grids.}
    \label{fig:multigrid}
\end{figure}

See Fig. \ref{fig:multigrid} for a low-resolution example of how the multigrid is implemented for this study. The benefit of this setup is that it provides increased resolution towards the inner disk where the orbital velocities changes over very small spatial scales. Also, it significantly shortens processing speeds when it comes to operations that need to be applied over the entire disk --- like seeing convolution. In the example, there are 3 concentric 8$\times$8 grids with undefined bounds x/y bounds, with each grid being 2 times shorter on the side (i.e. 4 times smaller) than the outer grid. Our models are constructed of 15 192$\times$192 grids with the bounds being x = y = 864.0~au. The innermost grid therefore has a resolution $2^{14}=16384$ times higher than the outermost grid.

\end{document}